\documentclass[12pt,a4paper]{article}
\usepackage{amsmath,amssymb}
\usepackage[dvips]{lscape, graphicx}

\DeclareMathOperator{\rre}{Re}

\begin{document}

\author{R.B.Nevzorov and M.A.Trusov}

\title{Particle spectrum in the modified NMSSM in the strong Yukawa coupling limit}

\maketitle

\begin{abstract}

\noindent A theoretical analysis of solutions of renormalisation
group equations in the MSSM corresponding to the quasi-fixed point
conditions shows that the mass of the lightest Higgs boson in this
case does not exceed $94\pm 5\text{~GeV}$. It means that a
substantial part of the parameter space of the MSSM is practically
excluded by existing experimental data from LEP\,II. In the NMSSM
the upper bound on the lightest Higgs boson mass reaches its
maximum in the strong Yukawa coupling regime, when Yukawa
constants are considerably larger the gauge ones on the Grand
Unification scale. In this paper a particle spectrum in a simple
modification of NMSSM which leads to a self-consistent solution in
the considered region of the parameter space is studied. This
model allows one to get $m_h\sim 125\text{~GeV}$ even for
comparatively low values of $\tan\beta\ge 1.9$. For an analysis of
the Higgs boson spectrum and neutralino spectrum a method for
diagonalisation of mass matrices proposed formerly is used. The
mass of the lightest Higgs boson in this model does not exceed
$130.5\pm 3.5\text{~GeV}$.

\end{abstract}

\newpage

\section{Introduction}

The search for the Higgs boson remains one of the top priorities
for existing accelerators as well as for those still at the design
stage. This is because this boson plays a key role in the Standard
Model which describes all currently available experimental data
with a high degree of accuracy. As a result of the spontaneous
symmetry breaking $SU(2)\otimes U(1)$ the Higgs scalar acquires a
nonzero vacuum expectation value without destroying the Lorentz
invariance, and generates the masses of all fermions and vector
bosons. An analysis of the experimental data using the Standard
Model has shown that there is a 95\% probability that its mass
will not exceed 210~GeV \cite{1}. At the same time, assuming that
there are no new fields and interactions and also no Landau pole
in the solution of the renormalisation group equations for the
self-action constant of Higgs fields up to the scale
$M_{\text{Pl}}\approx2.4\times 10^{18}\text{~GeV}$, we can show
that $m_h<180\text{~GeV}$ \cite{2}, \cite{3}. In this case,
physical vacuum is only stable provided that the mass of the Higgs
boson is greater than 135~GeV \cite{2}-\cite{6}. However, it
should be noted that this simplified model does not lead to
unification of the gauge constants \cite{7} and a solution of the
hierarchy problem \cite{8}. As a result, the construction of a
realistic theory which combines all the fields and interactions is
extremely difficult in this case.

Unification of the gauge constants occurs naturally on the  scale
$M_X\approx 3\times 10^{16}\text{~GeV}$ within the supersymmetric
generalisation of the Standard Model, i.e., the Minimal
Supersymmetric Standard Model (MSSM) \cite{7}. In order that all
the fundamental fermions acquire mass in the MSSM, not one but two
Higgs doublets $H_1$ and $H_2$ must be introduced in the theory,
each acquiring the nonzero vacuum expectation value $v_1$ and
$v_2$ where $v^2=v_1^2+v_2^2=(246\text{~GeV})^2$. The spectrum of
the Higgs sector of the MSSM contains four massive states: two
CP--even, one CP--odd, and one charged. An important
distinguishing feature of the supersymmetric model is the existing
of a light Higgs boson in the CP--even sector. The upper bound on
its mass is determined to a considerable extent by the value
$\tan\beta=v_2/v_1$. In the tree-level approximation the mass of
the lightest Higgs boson in the MSSM does not exceed the mass of
the Z-boson ($M_Z\approx 91.2\text{~GeV}$): $m_h\le M_Z|\cos
2\beta|$ \cite{9}. Allowance for the contribution of loop
corrections to the effective interaction potential of the Higgs
fields from a $t$--quark and its superpartners significantly
raises the upper bound on its mass:
\begin{equation}
m_h\le\sqrt{M_Z^2\cos^2
2\beta+\Delta^{(1)}_{11}+\Delta^{(2)}_{11}}\, . \label{1}
\end{equation}
Here $\Delta^{(1)}_{11}$ and $\Delta^{(2)}_{11}$ are the one--loop
\cite{10} and two--loop \cite{11} corrections, respectively. The
values of these corrections are proportional to $m_t^4$, where
$m_t$ is the running mass of $t$--quark which depends
logarithmically on the supersymmetry breaking scale $M_S$ and is
almost independent of the choice of $\tan\beta$. In \cite{3},
\cite{5}, \cite{6} bounds on the mass of the Higgs boson were
compared in the Minimal Standard and Supersymmetric models. The
upper bound on the mass of the light CP--even Higgs boson in the
MSSM increases with increasing $\tan\beta$ and for $\tan\beta\gg
1$ in realistic supersymmetric models with $M_S\le
1000\text{~GeV}$ reaches 125-128~GeV.

However, a considerable fraction of the solutions of the system of
MSSM renormalisation group equations is focused near the infrared
quasi-fixed point at $\tan\beta\sim 1$. In the region of parameter
space of interest to us ($\tan\beta\ll 50$) the Yukawa constants
of a $b$--quark ($h_b$) and a $\tau$--lepton ($h_\tau$) are
negligible so that an exact analytic solution can be obtained for
the one--loop renormalisation group equations \cite{12}. For the
Yukawa constants of a $t$--quark $h_t(t)$ and the gauge constants
$g_i(t)$ its solution has the following form:
\begin{gather}
Y_t(t)=\frac{\dfrac{E(t)}{6F(t)}}{1+\dfrac{1}{6Y_t(0)F(t)}},\quad
\tilde{\alpha}_i(t)=\frac{\tilde{\alpha}_i(0)}{1+b_i\tilde{\alpha}_i(0)t},\nonumber
\\
E(t)=\left[\frac{\tilde{\alpha}_3(t)}{\tilde{\alpha}_3(0)}\right]^{16/9}
\left[\frac{\tilde{\alpha}_2(t)}{\tilde{\alpha}_2(0)}\right]^{-3}
\left[\frac{\tilde{\alpha}_1(t)}{\tilde{\alpha}_1(0)}\right]^{-13/89}
,\quad F(t)=\int\limits_0^t E(t')dt', \label{2}
\end{gather}
where the index $i$ has values between 1 and 3,
\begin{gather*}
b_1=33/5,\quad b_2=1,\quad b_3=-3\\
\tilde\alpha_i(t)=\left(\frac{g_i(t)}{4\pi}\right)^2,\quad
Y_i(t)=\left(\frac{h_t(t)}{4\pi}\right)^2.
\end{gather*}
The variable $t$ is determined by a standard method
$t=\ln(M_X^2/q^2)$. The boundary conditions for the
renormalisation group equations are usually set at the grand
unification scale $M_X$ ($t=0$) where the values of all three
Yukawa constants are the same:
$\tilde\alpha_1(0)=\tilde\alpha_2(0)=\tilde\alpha_3(0)=\tilde\alpha(0).$
On the electroweak scale where $h_t^2(0)\gg 1$ the second term in
the denominator of the expression describing the evolution of
$Y_t(t)$ is much smaller than unity and all the solutions are
concentrated in a narrow interval near the quasi-fixed point
$Y_{\text{QFP}}(t)=E(t)/6F(t)$ \cite{13}. In other words in the
low-energy range the dependence of $Y_t(t)$ on the initial
conditions on the scale $M_X$ disappears. In addition to the
Yukawa constant of the $t$--quark, the corresponding trilinear
interaction constant of the scalar fields $A_t$ and the
combination of the scalar masses
$\mathfrak{M}^2_t=m_Q^2+m_U^2+m_2^2$ also cease to depend on
$A_t(0)$ and $\mathfrak{M}_t^2(0)$ as $Y_t(0)$ increases. Then on
the electroweak scale near the infrared quasi-fixed point $A_t(t)$
and $\mathfrak{M}_t^2(t)$ are only expressed in terms of the
gaugino mass on the Grand Unification scale. Formally this type of
solution can be obtained if $Y_t(0)$ is made to go to infinity.
Deviations from this solution are determined by ratio
$1/6F(t)Y_t(0)$ which is of the order of $1/10h_t^2(0)$ on the
electroweak scale.

The properties of the solutions of the system of MSSM
renormalisation group equations and also the particle spectrum
near the infrared quasi-fixed point for $\tan\beta\sim 1$ have
been studied by many authors \cite{14}, \cite{15}. Recent
investigations \cite{15}-\cite{17} have shown that for solutions
$Y_t(t)$ corresponding to the quasi-fixed point regime the value
of $\tan\beta$ is between 1.3 and 1.8. These comparatively low
values of $\tan\beta$ yield significantly more stringent bounds on
the mass of the lightest Higgs boson. The weak dependence of the
soft supersymmetry breaking parameters $A_t(t)$ and
$\mathfrak{M}_t^2(t)$ on the boundary conditions near the
quasi-fixed point means that the upper bound on its mass can be
calculated fairly accurately. A theoretical analysis made in
\cite{15}, \cite{16} showed that $m_h$ does not exceed $94\pm
5\text{~GeV}$. This bound is 25-30~GeV below the absolute upper
bound in the Minimal Supersymmetric Model. Since the lower bound
on the mass of the Higgs boson from LEP\,II data is 113~GeV
\cite{1}, which for the spectrum of heavy supersymmetric particles
is the same as the corresponding bound on the mass of the Higgs
boson in the Standard Model, a considerable fraction of the
solutions which come out to a quasi-fixed point in the MSSM, are
almost eliminated by existing experimental data. This provides the
stimulus for theoretical analyses of the Higgs sector in more
complex supersymmetric models.

The simplest expansion of the MSSM which can conserve the
unification of the gauge constants and raise the upper bound on
the mass of the lightest Higgs boson is the Nonminimal
Supersymmetric Standard Model (NMSSM) \cite{18}-\cite{20}. In
addition to the doublets $H_1$ and $H_2$, the Higgs sector of this
model contains the additional singlet superfield $Y$ relative to
the gauge $SU(2)\otimes U(1)$ interactions. The most attractive
region of the NMSSM parameter space from the point of view of
theoretical analysis is that corresponding to the limit of strong
Yukawa coupling when the Yukawa constants on the Grand Unification
scale $M_X$ are substantially larger than the gauge constant
$g_{\text{GUT}}$. This is the region where the upper bound on the
mass of the lightest Higgs boson reaches its maximum, which is
several GeVs larger than the corresponding absolute bound in the
MSSM. In addition, in this particular case it is possible to
select the interaction constants so as to achieve the unification
of the Yukawa constants of a $b$--quark and a $\tau$--lepton on
the scale $M_X$ \cite{21}, \cite{22} which usually occurs in GUTs
\cite{23}.

However, the mass of the lightest Higgs boson in the NMSSM differs
substantially from its upper bound \cite{24}. In this connection,
the present paper examines a very simple model in which $m_h$
reaches its upper theoretical bound for a specific choice of
fundamental parameters. This bound model is obtained by modifying
the Nonminimal Supersymmetric Model and yields a self-consistent
solution in the strong Yukawa coupling regime where, even for
comparatively low values of $\tan\beta\le 1.9$, the mass of the
lightest Higgs boson in the modified NMSSM may reach 125-127~GeV.
Although the parameter space of this model is enlarged
considerably, the theory does not lose its predictive capacity.
The proposed model is used to study characteristics of the
spectrum of superpartners of observable particles and Higgs
bosons. The mass of the lightest Higgs boson in this model does
not exceed $130.5\pm 3.5\text{~GeV}$.

This bound on the mass of the lightest Higgs boson is not the
absolute upper bound in supersymmetric models. For instance, it
was shown in \cite{25} that by introduction four or five
additional $5+\bar{5}$ multiplets of matter, the upper bound  on
$m_h$ in the NMSSM is increased up to 155~GeV. Recently the upper
bound on the mass of the lightest Higgs boson has been actively
discussed using more complex expansion of MSSM \cite{26},
\cite{28}. In particular, in addition to the singlet it is also
possible to introduce several $SU(2)$ triplets into the Higgs
sector of supersymmetric models. Their appearance destroys the
unification of the gauge constants at high energies. In order to
reconstruct this, in addition to triplets we also need to add
several multiplets of matter which carry colour charge in the
$SU(3)$ group but do not participate in $SU(2)\otimes U(1)$
interactions, for example four $3+\bar{3}$. A numerical analysis
made in \cite{27} shows that unification of the gauge constants
then occurs on the scale $\tilde{M}_X\sim 10^{17}\text{~GeV}$ and
the mass of the lightest Higgs boson does not exceed 190~GeV. The
existence of a fourth generation of particles in the MSSM
\cite{28}, which is extremely problematical from the point of view
of the known experimental data, also leads to an appreciable
increase in the upper bound on $m_h$. Consequently, an increase in
the upper bound on the mass of the lightest Higgs boson in the
supersymmetric models is usually accompanied by a substantial
increase in the number of particles in the models which may be
counted as a serious disadvantage of this type of model. In the
present study, unlike those noted above \cite{25}, \cite{28}, we
examine the dependence of $m_h$ and the particle spectrum on the
fundamental parameters of the modified NMSSM in the strong Yukawa
coupling regime.

\section{NMSSM parameters and their renormalisation in the strong Yukawa coupling regime}

By definition the superpotential of the Nonminimal Supersymmetric
Model is invariant with respect to the discrete transformations
$y'_\alpha=e^{2\pi i/3}y_\alpha$ of the $Z_3$ group \cite{19}
which means that we can avoid the problem of the $\mu$-term in
supergravity models. $Z_3$-symmetry usually occurs in string
models in which all the fields of the observable sector remain
massless in the exact supersymmetric limit. In addition to
observable superfields $y_\alpha$, supergravity theories also
contain a hidden sector in which local supersymmetry is broken. In
modern supergravity theories this sector includes singlet dilaton
$S$ and moduli $T_m$ fields with respect to gauge interactions.
These fields always appear in four-dimensional theory and they
occur as a result of the compactification of additional
dimensions. The vacuum-averaged dilaton and moduli fields
determine the values of the gauge constants on the Grand
Unification scale and also the dimensions and shape of compacted
space. The superpotential in supergravity models is usually
represented as an expansion in terms of superfields of the
observable sector \cite{29}:
\begin{equation}
W=\hat W_0(S,T_m)+\frac{1}{2}\mu_{\alpha\beta}(S,T_m)y_\alpha
y_\beta+\frac{1}{6}h_{\alpha\beta\gamma}(S,T_m)y_\alpha y_\beta
y_\gamma , \label{3}
\end{equation}
where $\hat W_0(S,T_m)$ is the superpotential of the hidden
sector. In expression (\ref{3}) summation is performed over the
recurrent greek subscripts. The requirements for conservation of
$R$-parity \cite{20} and gauge invariance have the result that the
single parameter $\mu$ is retained in the MSSM which corresponds
to the term $\mu(H_1,H_2)$ in the superpotential (\ref{3}).
However, the expansion (\ref{3}) assumes that this fundamental
parameter should be of the order of $M_{\text{Pl}}$ since this
scale is the only dimensional parameter characterising the hidden
(gravity) sector of the theory. In this case, however, the Higgs
bosons $H_1$ and $H_2$ acquire an enormous mass $m^2_{H_1,H_2}\sim
\mu^2\sim M^2_{\text{Pl}}$ and no breaking of $SU(2)\otimes U(1)$
symmetry occurs. In the NMSSM the term $\mu$ in the superpotential
(\ref{3}) is not invariant with respect to discrete
transformations of the $Z_3$ group and for this reason should be
eliminated from the analysis ($\mu=0$). As a result of the
multiplicative nature of the renormalisation of this parameter,
the term $\mu(q)$ remains zero on any scale $q\le M_X\div
M_{\text{Pl}}$. However, the absence of mixing of the Higgs
doublets on electroweak scale has the result that $H_1$ acquires
no vacuum expectation value as a result of the spontaneous
symmetry breaking and $d$--type quarks and charged leptons remain
massless. In order to ensure that all quarks and charged leptons
acquire nonzero masses, an additional singlet superfield $Y$ with
respect to gauge $SU(2)\otimes U(1)$ transformations is introduced
in the NMSSM. The superpotential of the Higgs sector of the
Nonminimal Supersymmetric Model \cite{18}-\cite{20} has the
following form:
\begin{equation}
W_h=\lambda Y(H_1H_2)+\frac{\varkappa}{3}Y^3. \label{4}
\end{equation}
As a result of the spontaneous breaking of $SU(2)\otimes U(1)$
symmetry, the field $Y$ acquires a vacuum expectation value
($\langle Y\rangle=y/\sqrt{2}$) and the effective $\mu$-term
($\mu=\lambda y/\sqrt{2}$) is generated.

In addition to the Yukawa constants $\lambda$ and $\varkappa$, and
also the Standard Model constants, the Nonminimal Supersymmetric
Model contains a large number of unknown parameters. These are the
so-called soft supersymmetry breaking parameters which are
required to obtain an acceptable spectrum of superpartners of
observable particles form the phenomenological point of view. The
hypothesis on the universal nature of these constants on the Grand
Unification scale allows us to reduce their number in the NMSSM to
three: the mass of all the scalar particles $m_0$, the gaugino
mass $M_{1/2}$, and the trilinear interaction constant of the
scalar fields $A$. In order to avoid strong violation of
CP--parity and also spontaneous breaking of gauge symmetry at high
energies ($M_{\text{Pl}}\gg E\gg m_t$) as a result of which the
scalar superpartners of leptons and quarks would require nonzero
vacuum expectation values, the complex phases of the soft
supersymmetry breaking parameters are assumed to be zero and only
positive values of $m_0^2$ are considered. Naturally universal
supersymmetry breaking parameters appear in the minimal
supergravity model \cite{31} and also in various string models
\cite{29}, \cite{32}. In the low-energy region the hypothesis of
universal fundamental parameters allows to avoid the appearance of
neutral currents with flavour changes and can simplify the
analysis of the particle spectrum as far as possible. The
fundamental parameters thus determined on the Grand Unification
scale should be considered as boundary conditions for the system
of renormalisation group equations which describes the evolution
of these constants as far as the electroweak scale or the
supersymmetry breaking scale. The complete system of the
renormalisation group equations of the Nonminimal Supersymmetric
Model can be found in \cite{33}, \cite{34}. These experimental
data impose various constraints on the NMSSM parameter space which
were analysed in \cite{35}, \cite{36}.

The introduction of the neutral field $Y$  in the NMSSM potential
leads to the appearance of a corresponding $F$-term in the
interaction potential of the Higgs fields. As a consequence, the
upper bound on the mass of the lightest Higgs boson is increased:
\begin{equation}
m_h\le\sqrt{\frac{\lambda^2}{2}v^2\sin^2 2\beta+M_Z^2\cos^2
2\beta+\Delta^{(1)}_{11}+\Delta^{(2)}_{11}} . \label{5}
\end{equation}
The relationship (\ref{5}) was obtained in the tree-level
approximation ($\Delta_{11}=0$) in \cite{20}. However, loop
corrections to the effective interaction potential of the Higgs
fields from the $t$--quark and its superpartners play a very
significant role. In terms of absolute value their contribution to
the upper bound on the mass of the Higgs boson remains
approximately the same as in the Minimal Supersymmetric Model.
When calculating the corrections $\Delta_{11}^{(1)}$ and
$\Delta_{11}^{(2)}$ we need to replace the parameter $\mu$ by
$\lambda y/\sqrt{2}$. Studies of the Higgs sector in the
Nonminimal Supersymmetric model and the one--loop corrections to
it were reported in \cite{24}, \cite{33}, \cite{36}-\cite{39}. In
\cite{6} the upper bound on the mass of the lightest Higgs boson
in the NMSSM was compared with the corresponding bounds on $m_h$
in the Minimal Standard and Supersymmetric Models. The possibility
of a spontaneous loss of CP--parity in the Higgs sector of the
NMSSM was studied in \cite{39}, \cite{40}.

It follows from condition (\ref{5}) that the upper bound on $m_h$
increases as $\lambda$ increases. Moreover, it only differs
substantially from the corresponding bound in the MSSM in the
range of small $\tan\beta$. For high values ($\tan\beta\gg 1$) the
value of $\sin 2\beta$ tends to zero and the upper bounds on the
mass of the lightest Higgs boson in the MSSM and NMSSM are almost
the same. The case of small $\tan\beta$ is only achieved for
fairly high values of the Yukawa constant of a $t$--quark $h_t$ on
the electroweak scale ($h_t(t_0)\ge 1$ where
$t_0=\ln(M_X^2/m_t^2)$), and $\tan\beta$ decreases with increasing
$h_t(t_0)$. However, an analysis of the renormalisation group
equations in the NMSSM shows that an increase of the Yukawa
constants on the electroweak scale is accompanied by an increase
of $h_t(0)$ and $\lambda(0)$ on the Grand Unification scale. It
thus becomes obvious that the upper bound on the mass of the
lightest Higgs boson in the Nonminimal Supersymmetric model
reaches its maximum on the strong Yukawa coupling limit, i.e.,
when $h_t(0)\gg g_i(0)$ and $\lambda(0)\gg g_i(0)$.

In our previous two studies \cite{21}, \cite{41} we analysed the
renormalisation of the NMSSM parameters in the strong Yukawa
coupling regime. We showed \cite{21} that as the values of the
Yukawa constants on the scale $M_X$ increase, the solutions of the
renormalisation group equations on the electroweak scale are
pulled towards a quasi-fixed (Hill) line ($\varkappa=0$) or
surface ($\varkappa\ne 0$) in Yukawa constant space, which limit
the range of permissible values of $h_t$, $\lambda$ and
$\varkappa$. Outside this range in the solutions of the
renormalisation group equations for $Y_i(t)$, where
$Y_t(t)=h_t^2/(4\pi)^2$, $Y_\lambda(t)=\lambda^2/(4\pi)^2$, and
$Y_\varkappa(t)=\varkappa^2/(4\pi)^2$, a Landau pole appears below
the Grand Unification scale and perturbation theory can not be
applied when $q^2\sim M_X^2$. Along the Hill line or surface the
values of $Y_i(t)$ are distributed nonuniformly. As $Y_i(0)$
increases , the region in which the solutions of the
renormalisation group equations are concentrated on the
electroweak scale in the strong Yukawa coupling regime becomes
narrower and in the limit $Y_i(0)\to\infty$ all the solutions are
focused near the quasi-fixed points. These points are formed as a
result of intersection of the Hill line or surface with the
infrared fixed (invariant) line. This line connects the stable
fixed point in the strong Yukawa coupling regime \cite{42} with
the infrared stable fixed point of the system of NMSSM
renormalisation group equations \cite{43}. The invariant lines
their properties in the Minimal Standard and Supersymmetric Models
were studied in detail in \cite{44}.

As with increasing $Y_t(0)$ the Yukawa constants approach the
quasi-fixed points, corresponding solutions for the trilinear
constants $A_i(t)$ and the combinations of scalar particle masses
$\mathfrak{M}^2_i(t)$
\begin{align*}
\mathfrak{M}^2_t&=m_Q^2+m_U^2+m_2^2,\\
\mathfrak{M}_\lambda^2&=m_2^2+m_1^2+m_y^2,\\
\mathfrak{M}^2_\varkappa&=3 m_y^2,
\end{align*}
cease to depend on their initial values on the scale $M_X$. If the
evolution of the gauge and Yukawa constants is known, the rest of
the renormalisation group equations of the NMSSM can be considered
as a system of linear equations for the soft symmetry breaking
parameters. In order to solve this system of equations we first
need to integrate the equations for the gaugino masses and for the
trilinear interaction constants of the scalar fields $A_i(t)$ and
then use the results to calculate $\mathfrak{M}_i^2(t)$. Since the
system of differential equations for $A_i(t)$ and
$\mathfrak{M}_i^2(t)$ is linear, under universal boundary
conditions we can obtain the dependence of the soft supersymmetry
breaking parameters on the electroweak scale on $A$, $M_{1/2}$,
and $m_0^2$ \cite{45}, \cite{46}:
\begin{gather}
A_i(t)=e_i(t)A+f_i(t)M_{1/2}, \nonumber \\
\mathfrak{M}_i^2(t)=a_i(t)m_0^2+b_i(t)M_{1/2}^2+c_i(t)AM_{1/2}+d_i(t)A^2.
\label{6}
\end{gather}
The functions $e_i(t)$, $f_i(t)$, $a_i(t)$, $b_i(t)$, $c_i(t)$,
and $d_i(t)$ remain unknown since no analytic solution of the
complete system of NMSSM renormalisation group equations exists.
It was shown in \cite{41} that as the quasi-fixed points are
approached, the values of the functions $e_i(t_0)$, $a_i(t_0)$,
$c_i(t_0)$, and $d_i(t_0)$ tend to zero whereas for
$Y_i(0)\to\infty$ all $A_i(t)$ are proportional to $M_{1/2}$ and
all $\mathfrak{M}_i^2(t)\propto M_{1/2}^2$. the weak dependence of
$A_i(t)$ and $\mathfrak{M}_i^2(t)$ in the strong Yukawa coupling
regime on the initial conditions has the result that the solutions
of the renormalisation group equations for trilinear interaction
constants and combinations of scalar particle masses and also the
solutions for $Y_i(t)$ are focused on the electroweak scale near
the quasi-fixed points. In general under nonuniversal boundary
conditions the solutions for $A_i(t)$ and $\mathfrak{M}^2_i(t)$
are grouped near certain lines ($\varkappa=0$) or planes
($\varkappa\ne 0$) in the soft supersymmetry breaking parameter
space. These lines and planes are almost perpendicular to the axes
$A_t$ and $\mathfrak{M}_t^2$ whereas the planes in the spaces
$(A_t.A_\lambda,A_\varkappa)$ and
$(\mathfrak{M}_t^2,\mathfrak{M}_\lambda^2,\mathfrak{M}_\varkappa^2)$
are also almost parallel to the axes $A_\varkappa$ and
$\mathfrak{M}_\varkappa^2$. Along these lines and planes as
$Y_i(0)$ increases, the trilinear interaction constants and
combinations of scalar particle masses go to quasi-fixed points.

\section{Choice of model}

The soft supersymmetry breaking parameters play a key role in an
analysis of the particle spectrum in modern supersymmetric models.
They destroy the Bose--Fermi degeneracy of the spectrum in
supersymmetric theories so that the superpartners of observable
particles are substantially heavier than quarks and leptons.
However, it should be noted that a study of the particle spectrum
in the NMSSM is considerably more complex than a study of this
spectrum in the MSSM for $\tan\beta\sim 1$ since two new Yukawa
constants $\lambda$ and $\varkappa$ appear in the nonminimal
supersymmetric model for which the boundary conditions are
unknown. In turn, the renormalisation of the trilinear interaction
constants and the scalar particle masses, i.e, the values of the
functions $e_i(t_0)$, $f_i(t_0)$, $a_i(t_0)$, $b_i(t_0)$,
$c_i(t_0)$, and $d_i(t_0)$, where
$t_0=2\ln(M_X/M_t^{\text{pole}})$ depends on the choice of
$h_t(0)$, $\lambda(0)$, and $\varkappa(0)$ on the grand
unification scale.

The most interesting from the point of view of a theoretical
analysis is a study of the spectrum of heavy supersymmetric
particles when the scale of the supersymmetry breaking $M_S^2\gg
M_Z^2$. This is primarily because in this limit the contribution
of new particles to the electroweak observable ones is negligible
(see, for example \cite{47}). As has been noted, the Standard
Model highly accurately describes all the existing experimental
data. Additional Higgs fields and superpartners of observable
particles interacting with vector $W^\pm$ and $Z$ bosons make a
nonzero contribution to the electroweak observables. However, for
$M_S^2\gg M_Z^2$ their contribution is suppressed in a power
fashion as $(M_Z/M_S)^2$, where any increase in the scale of the
supersymmetry breaking leads to convergence of the theoretical
predictions for the strong interaction constant $\alpha_s(M_Z)$
which may be obtained assuming unification of the gauge constants
\cite{48}, with the results of an analysis of the experimental
data \cite{49}. In addition, it should be noted that the mass of
the lightest Higgs boson which is one of the central objects of
investigation in any supersymmetric model reaches its highest
value for $M_S\sim 1-3\text{~TeV}$.

Unfortunately, in the strong Yukawa coupling regime in the NMSSM
with a minimal set of fundamental parameters it is impossible to
obtain a self-consistent solution which on the one hand would lead
to a spectrum with heavy superparticles and on the other could
give a mass of the lightest Higgs boson greater than that in the
MSSM. When calculating the particle spectrum, the fundamental
parameters $A$, $m_0^2$, and $M_{1/2}$ on the scale $M_X$ should
be selected so that the derivatives of the interaction potential
of the scalar potential of the scalar fields $V(H_1,H_2,Y)$ with
respect to the vacuum expectation values $v_1$, $v_2$, and $y$
would be zero at the minimum:
\begin{equation}
\frac{\partial V(v_1,v_2,y)}{\partial v_1}=0,\quad \frac{\partial
V(v_1,v_2,y)}{\partial v_2}=0,\quad \frac{\partial
V(v_1,v_2,y)}{\partial y}=0. \label{7}
\end{equation}
Since the trilinear interaction constants and the scalar particle
masses in the strong Yukawa coupling regime are almost independent
of $A$, Eqs.(\ref{7}) link the vacuum expectation value of the
neutral scalar field $\langle Y\rangle$, and the parameters
$m_0^2$ and $M_{1/2}$. The value of $\tan\beta$ is determined
using the Yukawa constant of a $t$--quark on the electroweak scale
(see below). Then a spectrum of heavy supersymmetric particles is
only achieved when $\lambda/\varkappa\gg 1$. However, in this
region of parameter space the value of $m_h^2$ becomes negative
and the physical vacuum is unstable which can be attributed to the
strong mixing of the CP--even components of the neutral field $Y$
and the superposition of Higgs doublets
$h=H_1\cos\beta+H_2\sin\beta$.

Studies of the particle spectrum in the nonminimal supersymmetric
model \cite{33},\cite{45},\cite{46},\cite{50} have shown that a
self-consistent nontrivial solution of the system of nonlinear
algebraic Eqs.(\ref{7}) for $|\langle Y\rangle|\le 10\text{~TeV}$
which determines the position of the minimum of the interaction
potential of the scalar fields, only exists for
$\lambda^2(t_0),\varkappa^2(t_0)\lesssim 0.1$. In this case a
strict correlation exists between the fundamental parameters of
the NMSSM. In particular, in order to ensure that spontaneous
breaking of $SU(2)\otimes U(1)$ symmetry occurs and the field $Y$
has a nonzero vacuum expectation value of the field, the condition
$|A_\varkappa/m_y|\ge 3$ must be satisfied. However, the following
inequalities must also be satisfied:
\begin{align*}
A_l^2&\le 3(m_1^2+m_{E_L}^2+m_{E_R}^2),\\ A_d^2&\le
3(m_1^2+m_{D_L}^2+m_{D_R}^2),\\ A_u^2&\le
3(m_2^2+m_{U_L}^2+m_{U_R}^2).
\end{align*}
Otherwise, the superpartners of leptons and quarks acquire vacuum
expectation values \cite{51}. All these constraints have the
result that the ratio $|A/m_0|$ varies between 3 and 4. In
\cite{52}, \cite{53} the particle spectrum in the NMSSM is
analysed separately for $\tan\beta=m_t(m_t)/m_b(m_t)$ and under
nonuniversal boundary conditions.

The limit $h_t^2\gg\lambda^2,\varkappa^2$ in the nonminimal
supersymmetric model corresponds to the MSSM \cite{33}. For
$\varkappa=0$ the Lagrangian of the Higgs sector of the NMSSM is
invariant with respect to the global $SU(2)\otimes U(1)\otimes
U(1)$ transformations. As a result of the spontaneous symmetry
breaking, only the $U(1)$ symmetry corresponding to
electromagnetic interaction remains unbroken, which leads to four
massless degrees of freedom. Two of these are eaten by a charged
$W^\pm$ boson and one by a $Z$ boson. Ultimately, the spectrum of
the nonminimal supersymmetric model for $\varkappa=0$ contains one
physical massless state which corresponds to the CP--odd component
of the field $Y$. For low values $\varkappa^2\ll\lambda^2 h_t^2$
the mass of the lightest CP--odd boson is nonzero and is
proportional to the self-action constant of the neutral superfield
$Y$. If the Yukawa constants $\lambda,\varkappa\sim
10^{-3}-10^{-4}$, for a certain choice of fundamental parameters
the mass of the lightest CP--even Higgs boson may be only a few
GeV \cite{33},\cite{46},\cite{54}. The main contribution to its
wave function is made by the neutral scalar field $Y$ which makes
it very difficult to search for this on existing accelerators and
those at the design stage since the interaction constants of this
type of Higgs boson with gauge bosons and fermions are small. In
this limiting case, the lightest stable supersymmetric particle
having R--parity of $-1$ is usually the superpartner of the
neutral scalar field $Y$ \cite{33}, \cite{46}.

However, unlike the minimal supersymmetric model, the discrete
$Z_3$ symmetry which can avoid problems of the $\mu$--term in the
NMSSM has the result that three degenerate vacuums appear in the
theory because of the breaking of gauge symmetry. Immediately
after a phase transition on the electroweak scale the is filled
equally with three degenerate phases. The entire space is then
divided into separate regions in each of which a particular case
is achieved. The regions are separated by domain walls with the
surface energy density $\sigma\sim v^3$. Data from cosmological
observations eliminate the existence of domain walls. The domain
structure of vacuum in the NMSSM is destroyed if the vacuum
degeneracy \cite{55} caused by $Z_3$ symmetry disappears. It was
shown in \cite{56} that breaking of $Z_3$ symmetry by introducing
into the NMSSM Lagrangian nonrenormalisable operators of dimension
$d=5$ which do not break the $SU(2)\otimes U(1)$ symmetry can be
used to obtain splitting of initially degenerate vacuums such that
the domain walls disappear before the beginning of the
nucleosynthesis are ($T\sim 1\text{~MeV}$). Although operators of
dimension $d=5$ are suppressed with respect to $M_{\text{Pl}}$ in
supergravity models, their introduction leads to quadratic
divergences in the two--loop approximation, i.e., to the problem
of hierarchies. Consequently, linear and quadratic terms with
respect to superfields are generated in the superpotential of this
theory and the vacuum expectation value of the neutral scalar $Y$
is of the order of $10^{11}\text{~GeV}$.

In order to avoid a vacuum domain structure and obtain a
self-consistent solution in the strong Yukawa coupling regime, we
need to modify the nonminimal supersymmetric model. The NMSSM can
be modified most simply by introducing additional terms in the
superpotential of the Higgs sector: $\mu(H_1 H_2)$ and $\mu'Y^2$
which are not forbidden by gauge $SU(2)\otimes U(1)$ and R
symmetries. The additional bilinear terms in the NMSSM
superpotential destroy the $Z_3$ symmetry and no domain structures
appear in this theory since no system of degenerate vacuums
exists. The introduction of the parameter $\mu$ ensures that it is
possible to obtain a spectrum of heavy supersymmetric particles in
the strong Yukawa coupling regime in the modified model and for a
certain choice of $\mu'$ the mass of the lightest Higgs boson
reaches its upper bound. In this case the mass of the lightest
Higgs boson has its highest value $\varkappa=0$ since as
$\varkappa(t_0)$ increases, the upper bound on its mass is reduced
as a result of a decrease in $\lambda(t_0)$. In the limit
$\varkappa=0$ the CP--odd Higgs sector of the modified NMSSM
contains no physical massless states since in this case, the
global symmetry of the Lagrangian is the same as the local
symmetry which eliminates the Yukawa self-action constant of the
neutral field $Y$ from the analysis. Assuming that this is zero
and neglecting all the Yukawa constants except for $\lambda$ and
$h_t$, the complete superpotential of the modified NMSSM can be
expressed in the following form:
\begin{equation}
W_{\text{NMSSM}}=\mu(H_1H_2)+\mu'Y^2+\lambda
Y(H_1H_2)+h_t(H_2Q)U_R^C\, . \label{8}
\end{equation}
where $U_R^C$ is the charge-coupled right superfield of a
$t$--quark and $Q$ is a doublet of left superfields of $b$ and
$t$--quarks.

In supergravity models, bilinear terms with respect to the
superfields may be generated in the superpotential (\ref{8}) as a
result of the additional term $Z(H_1H_2)+h.c.$ in the K\"ahler
potential \cite{57},\cite{58} or nonrenormalisable interaction of
the fields of the observable and hidden sectors. The appearance of
nonrenormalisable operators of this type in the superpotential of
supergravity models may be attributed to nonperturbative effects
(for instance, gaugino condensation) \cite{58},\cite{59}. In
addition to the parameters $\mu$ and $\mu'$, this model also sees
the appearance of the corresponding bilinear interaction constants
of the scalar fields $B$ and $B'$ which for a minimal choice of
fundamental parameters should be assumed to the equal on the grand
unification scale. Thus, the nonminimal supersymmetric model may
include seven fundamental parameters in addition to the constants
of the Standard Model:
\[
\lambda,\; \mu,\; \mu',\; A,\; B,\; m_0,\; M_{1/2}\, .
\]

\section{Constraints on the parameter space of the modified NMSSM}

Despite a substantial expansion of the parameter space, the theory
does not lose its predictive capacity. An analysis of the
behaviour of the solutions of the NMSSM renormalisation group
equations in the strong Yukawa coupling limit for $\varkappa=0$
showed that for $Y_i(0)\to\infty$ all the solutions are
concentrated near the quasi-fixed point:
\begin{equation}
\begin{array}{lll}
\rho_t^{\text{QFP}}(t_0)=0.803\, , &
\rho_{A_t}^{\text{QFP}}(t_0)=1.77\, , &
\rho_{\mathfrak{M}_t^2}^{\text{QFP}}(t_0)=6.09\\[3mm]
\rho_\lambda^{\text{QFP}}(t_0)=0.224\, , &
\rho_{A_\lambda}^{\text{QFP}}(t_0)=-0.42\, , &
\rho_{\mathfrak{M}_\lambda^2}^{\text{QFP}}(t_0)=-2.28\, ,
\end{array}
\label{9}
\end{equation}
where $\rho_t(t)=\dfrac{Y_t(t)}{\tilde{\alpha}_3(t)}$,
$\rho_\lambda(t)=\dfrac{Y_\lambda(t)}{\tilde{\alpha}_3(t)}$,
$\rho_{A_i}(t)=\dfrac{A_i(t)}{M_{1/2}}$ and
$\rho_{\mathfrak{M}_\lambda^2}(t)=\dfrac{\mathfrak{M}_\lambda^2(t)}{M_{1/2}^2}$.
Thus, at the first stage of the analysis we fixed the initial
values of the Yukawa constants $\lambda^2(0)=h_t^2(0)=10$
corresponding to the quasi-fixed point regime (\ref{9}) of the
renormalisation group equations and also the supersymmetry
breaking scale $M_3(1000\text{~GeV})=1000\text{~GeV}$ which
determines the mass scale of all the supersymmetric particles.

Existing FNAL experimental data from measurements of the mass of a
$t$--quark can uniquely relate $\tan\beta$ to the Yukawa constant
$h_t$ of a $t$--quark. The running mass of a $t$--quark generated
when the $SU(2)\otimes U(1)$ symmetry is broken is directly
proportional to $h_t(t_0)$:
\begin{equation}
m_t(M_t^{\text{pole}})=\frac{h_t(M_t^{\text{pole}})}{\sqrt{2}}v
\sin\beta. \label{10}
\end{equation}
However, the value of $m_t(M_t^{\text{pole}})$ calculated in the
$\bar{MS}$ scheme \cite{60} is equal to
$m_t(M_t^{\text{pole}})=165\pm 5\text{~GeV}$. The inaccuracy in
determining the running mass of a $t$--quark is primarily
attributable to the experimental error with which its pole mass is
measures ($M_t^{\text{pole}}=174.3\pm 5.1\text{~GeV}$ \cite{61}).
For each fixed set of boundary conditions $h_t(0)$ and
$\lambda(0)$, using renormalisation group equations we can
calculate the Yukawa constant of a $t$--quark on the electroweak
scale and then, substituting the value obtained $h_t(t_0)$ into
formula (\ref{10}), we can determine the value of $\tan\beta$. In
the infrared quasi-fixed point regime we obtain $\tan\beta\approx
1.88$ for $m_t(M_t^{\text{pole}})=165\text{~GeV}$ (Tables 1 and
2).

An additional constant which fixes $M_3$ can be used to determine
one of the supersymmetry breaking parameters $M_{1/2}$. The values
of all the other dimensional parameters $\mu$, $\mu'$, $A$, $B$,
and $m_0$ should be selected so that spontaneous breaking of gauge
$SU(2)\otimes U(1)$ symmetry occurs on the electroweak scale. The
complete interaction potential of the Higgs fields in the modified
NMSSM can be expressed as the sum:
\begin{multline}
V(H_1,H_2,Y)=\mu_1^2|H_1|^2+\mu_2^2|H_2|^2+\mu_y|Y|^2+\Bigl[\mu_3^2(H_1H_2)+\mu_4^2Y^2+
\lambda A_\lambda Y(H_1H_2)+{}\\
{}+\lambda\mu'Y^*(H_1H_2)+\lambda\mu Y
(|H_1|^2+|H_2|^2)+h.c.\Bigr]+\lambda^2|(H_1H_2)|^2+\lambda^2Y^2(|H_1|^2+|H_2|^2)+{}\\
{}+\frac{{g'}^2}{8}
(|H_2|^2-|H_1|^2)^2+\frac{g^2}{8}(H_1^+\boldsymbol{\sigma}H_1+H_2^+\boldsymbol{\sigma}H_2)^2+
\Delta V(H_1,H_2,Y)\, , \label{11}
\end{multline}
where $\Delta V(H_1,H_2,Y)$ are the one--loop corrections to the
effective interaction potential; $g$ and $g'$ are the constants of
the gauge $SU(2)$ and $U(1)$ interactions ($g_1=\sqrt{5/3}g'$).
The constants $\mu_i^2$ in the interaction potential (\ref{11})
are related to the soft supersymmetry breaking parameters as
follows:
\begin{gather*}
\mu_1^2=m_1^2+\mu^2\, ,\quad \mu_2^2=m_2^2+\mu^2\, ,\quad
\mu_y^2=m_y^2+{\mu'}^2\, ,\\ \mu_3^2=B\mu\, ,\quad
\mu_4^2=\frac{1}{2}B'\mu'\, .
\end{gather*}
where
\begin{gather*}
m_1^2(M_X)=m_2^2(M_X)=m_y^2(M_X)=m_0^2\, ,\\
B(M_X)=B'(M_X)=B_0\,
.
\end{gather*}
The position of the physical minimum of the interaction potential
of the Higgs fields (\ref{11}) is determined by Eqs.(\ref{7}).
Since the vacuum expectation value $v$ and $\tan\beta$ are known,
the system of Eqs.(\ref{7}) can be used to find $\mu$ and $B_0$.
Then, it is convenient to introduce
$\mu_{\text{eff}}=\mu+\dfrac{\lambda y}{\sqrt{2}}$ instead of
$\mu$. After various transformations we obtain:
\begin{gather}
\mu_{\text{eff}}^2=\frac{m_1^2-m_2^2\tan^2\beta+\Delta_Z(\mu_{\text{eff}})}
{\tan^2\beta-1}
 -\frac{1}{2}M_Z^2 \nonumber \\
\left(m_1^2+m_2^2+2\mu_{\text{eff}}^2+\frac{\lambda^2}{2}v^2
 +\Delta_\beta(\mu_{\text{eff}})\right)\sin 2\beta=-2\left(B\mu
 +\frac{\lambda y X_2}{\cos 2\beta}\right) \label{12} \\
y\left(m_y^2+{\mu'}^2+B'\mu'\right)=\frac{\lambda}{2}v^2X_1-
\Delta_y(\mu_{\text{eff}})\, , \nonumber
\end{gather}
where $\Delta_i$ corresponds to the contribution of the one--loop
corrections:
\begin{gather*}
\Delta_\beta=\frac{2}{v^2\tan 2\beta}\frac{\partial\Delta
V}{\partial\beta}+4\frac{\partial\Delta V}{\partial v^2}\, ,\quad
\Delta_y=\frac{\partial\Delta V}{\partial y}\, ,\\
\Delta_Z=\frac{1}{\cos^2\beta}\left\{ 2\frac{\partial V}{\partial
 v^2}\cos 2\beta-\frac{1}{v^2}\frac{\partial\Delta
 V}{\partial\beta}\sin 2\beta\right\}\, ,
\end{gather*}
and
\[
X_1=\frac{1}{\sqrt{2}}\left(2\mu_{\text{eff}}+(\mu'+A_{\lambda})
\sin 2\beta\right),\quad
X_2=\frac{1}{\sqrt{2}}\left(\mu'+A_{\lambda}\right)\cos 2\beta.
\]

When calculating the one--loop corrections we shall only take into
account the contribution from loops containing a $t$--quark and
its superpartners since their contribution is dominant. In
supersymmetric theories each fermion state with a specific
chirality has a scalar superpartner. Thus, a $t$--quark
incorporating left and right chiral components has two scalar
superpartners, right $\tilde{t}_R$ and left $\tilde{t}_L$, which
become mixed as a result of the spontaneous breaking of
$SU(2)\otimes U(1)$ symmetry, and this results in the formation of
two charged scalar particles having masses $m_{\tilde{t}_1}$ and
$m_{\tilde{t}_2}$:
\begin{equation}
m_{\tilde t_1, \tilde t_2}^2=\frac{1}{2}(m_Q^2+m_U^2+2m_t^2
\pm\sqrt{(m_Q^2-m_U^2)^2+4m_t^2X_t^2})\, , \label{13}
\end{equation}
where $X_t=A_t+\mu_{\text{eff}}/\tan\beta$. Since
$m_{\tilde{t}_1}^2$ and $m_{\tilde{t}_2}^2$ should be positive we
have $X_t^2<\dfrac{1}{m_t^2}(m_Q^2+m_t^2)(m_U^2+m_t^2)$.
Otherwise, the quark fields acquire nonzero vacuum expectation
values and the gauge $SU(2)\otimes U(1)$ symmetry of the initial
Lagrangian is completely broken, which leads to the appearance of
nonzero masses for gluons and photons. The contribution of the
one--loop corrections from the $t$--quark and its superpartners to
the effective interaction potential of the Higgs fields is
expressed only in terms of their masses:
\begin{multline}
\Delta V(H_1,H_2,Y)=\frac{3}{32\pi^2}\left(m_{\tilde
t_1}^4\left(\log\frac{m_{\tilde
t_1}^2}{q^2}-\frac{3}{2}\right)+{}\right.\\ \left.{}+m_{\tilde
t_2}^4\left(\log\frac{m_{\tilde
t_2}^2}{q^2}-\frac{3}{2}\right)-2m_t^4\left(\log\frac{m_t^2}{q^2}-\frac{3}{2}\right)\right)\,
.\label{14}
\end{multline}
For this reason all $\Delta_i$ are merely functions of
$\mu_{\text{eff}}$ and do not depend on $B_0$ and $y$.

Using the first equation of the system (\ref{12}) we can find
$\mu_{\text{eff}}$. In this case, the sign of $\mu_{\text{eff}}$
is not fixed and must be considered as a free parameter in the
theory. Substituting this value of $\mu_{\text{eff}}$ into the two
remaining equations of the system (\ref{12}), we can eliminate
$B_0$ from the number of independent fundamental parameters and
calculate the vacuum expectation value of the field $Y$: $\langle
Y\rangle=y/\sqrt{2}$ and in order to find $B_0$ we need to bear in
mind the relationships linking the bilinear interaction constants
$B$ and $B'$ in the electroweak scale with $B_0$ obtained by
solving the renormalisation group equations in the modified NMSSM
(see Appendix):
\begin{align}
B(t)&=\zeta(t)B_0+\sigma(t)A+\omega(t)M_{1/2},\nonumber \\
B'(t)&=B_0+\sigma_1(t)A\frac{\mu_0}{\mu_0'}+\omega_1(t)M_{1/2}
\frac{\mu_0}{\mu_0'}\, ,\label{15}
\end{align}
where $\zeta(t)$, $\sigma(t)$, $\sigma_1(t)$, $\omega(t)$, and
$\omega_1(t)$ are various functions of $t$, $h_t(0)$, and
$\lambda(0)$ which do not depend on the choice of fundamental soft
supersymmetry breaking parameters on the grand unification scale,
and also on $\mu_0$ and $\mu'_0$. For a fixed sign of
$\mu_{\text{eff}}$ three different solutions of the system of
Eqs.(\ref{12}) exist. However, only one of these is of interest
from the physical point of view. It follows from the last equation
in (\ref{12}) that the vacuum expectation value of the field $Y$
is of the order of $y\sim\lambda v^2/M_S$ and for the case of
heavy supersymmetric particles $y\ll v$. The other two solutions
give an excessively light CP--even Higgs boson which corresponds
to fine tuning between the fundamental parameters $B_0$ and
$\mu'$.

The parameters $\mu_{\text{eff}}$ and $B_0$ thus determined and
also the vacuum expectation value $y$ depend on the choice of $A$,
$m_0$, and $\mu'$. Thus, at the next stage of the analysis of the
modified NMSSM we studied the dependence of the particle spectrum
on these fundamental parameters using Eqs.(\ref{6}) linking
$A_i(t_0)$ and $\mathfrak{M}_i^2(t_0)$ to $A$ and $m_0^2$.
Similarly we investigated the spectrum of superpartners of
observable particles and Higgs bosons for other values of the
Yukawa constants from the vicinity of the infrared quasi-fixed
point. Although for $\tan\beta\le 2$ in the strong Yukawa coupling
regime the parameters $h_t$ and $\lambda$ can be selected so that
the Yukawa constants of a $b$--quark and a $\tau$--lepton would be
the same on the grand unification scale \cite{21}, \cite{22}, when
studying the particle spectrum in the modified NMSSM we do not
confine ourselves to the case $R_{b\tau}(0)=1$ where
$R_{b\tau}=h_b(t)/h_\tau(t)$. This condition arises in minimal
schemes of gauge interaction unification \cite{23} and imposes
very stringent constraints on the parameter space of the model
being studied. However, since $h_b$ and $h_\tau$ for
$\tan\beta\sim 1$ have small absolute values, they can be
generated by means of nonrenormalisable operators as a result of
the spontaneous symmetry breaking on the scale $M_X$ and in this
case, the Yukawa constants of a $b$--quark and a $\tau$--lepton
may differ.

\section{Calculations of masses of Higgs bosons and neutralinos}

We shall first consider the Higgs sector in the modified NMSSM
which includes three CP--even states, two CP--odd, and one charged
Higgs boson. The determinants of the mass matrices of the CP--odd
and charged Higgs bosons go to zero which corresponds to the
appearance of two Goldstone bosons which are eaten by massive
vector $W^\pm$ and $Z$ bosons during spontaneous breaking of
$SU(2)\otimes U(1)$ symmetry. The $(3\times 3)$ mass matrix of the
CP--odd sector is formed by mixing the imaginary parts of the
neutral components of the Higgs doublets with the imaginary part
of the field $Y$. However, since the determinant of this matrix is
zero, the problem of finding the eigenvalues reduces to solving an
ordinary quadratic equation. The calculated masses of the CP--odd
states in the modified NMSSM are
\begin{equation}
m_{A_1,A_2}^2=\frac{1}{2}\left(m_A^2+m_B^2\pm\sqrt{
\left(m_A^2-m_B^2\right)^2+ 4\left(\frac{\lambda v}{\sqrt 2}
(\mu'+A_\lambda)+\Delta_0 \right)^2}~\right), \label{16}
\end{equation}
\[
m_A^2=m_1^2+m_2^2+2\mu_{\text{eff}}^2+\frac{\lambda^2}{2}v^2+
\Delta_A,\quad m_B^2=m_y^2+{\mu'}^2-B'\mu'+\frac{\lambda^2}{2}v^2+
\Delta_3,
\]
where $\Delta_i$ are the one--loop corrections
\cite{33},\cite{36},\cite{38}.

A more complex situation is encountered in the sector of the
CP--even Higgs fields which appears as a result of mixing of the
real parts of the neutral components of the Higgs doublets and the
field $Y$. The determinant of the mass matrix of the CP--even
sector ix nonzero and thus in order to calculate its eigenvalues
we need to solve a cubic equation. However, for the case of heavy
supersymmetric particles ($M_S\gg M_Z$) in the Higgs field basis
\begin{eqnarray}
\chi_1&=&\frac{1}{\sqrt{2}}\cos\beta \rre H_1^0+
\frac{1}{\sqrt{2}}\sin\beta \rre H_2^0 \nonumber \\
\chi_2&=&-\frac{1}{\sqrt{2}}\sin\beta \rre H_1^0+
\frac{1}{\sqrt{2}}\cos\beta \rre H_2^0 \label{17} \\
\chi_3&=&\frac{1}{\sqrt{2}} \rre Y \nonumber
\end{eqnarray}
this matrix has a hierarchical structure:
\begin{equation}
M_{ij}^2=\begin{pmatrix}
E_1^2&0&0\\0&E_2^2&0\\0&0&E_3^2
\end{pmatrix}+\begin{pmatrix}
V_{11}&V_{12}&V_{13}\\V_{21}&V_{22}&V_{23}\\V_{31}&V_{32}&V_{33}
\end{pmatrix} \label{18}
\end{equation}
where
\begin{gather*}
E_1^2=0,\qquad E_2^2=m_1^2+m_2^2+2\mu_{\text{eff}}^2,\qquad
E_3^2=m_y^2+{\mu'}^2+B'\mu'\, ,\\ V_{11}=M_Z^2\cos^2
2\beta+\frac{1}{2}\lambda^2 v^2\sin^2 2\beta+ \Delta_{11},\qquad
V_{13}=V_{31}=\lambda vX_1+\Delta_{13}\, ,\\
V_{12}=V_{21}=\left(\frac{1}{4}\lambda^2
v^2-\frac{1}{2}M_Z^2\right)\sin 4\beta+\Delta_{12},\qquad
V_{23}=V_{32}=\lambda vX_2+\Delta_{23}\, ,\\ V_{22}=M_Z^2\sin^2
2\beta+\frac{1}{2}\lambda^2 v^2 \cos^2
2\beta+\Delta_A+\Delta_{22},\qquad V_{33}=\frac{1}{2}\lambda^2
v^2+\Delta_{33}\, .
\end{gather*}
Here $\Delta_A$ and $\Delta_{ij}$ are the corrections from loops
containing the $t$--quark and its superpartners. The hierarchical
structure of the mass matrix means that perturbation from quantum
mechanics can be used to diagonalise it. The role of the smallness
parameters in the perturbation theory are played by the ratios
$M_Z^2/E_2^2$ and $M_Z^2/E_3^2$. This method of calculating the
masses of Higgs bosons in supersymmetric theories was developed in
\cite{24}. Also discussed there is the simplest method of
obtaining a hierarchical mass matrix in the Higgs field basis
(\ref{17}). A numerical analysis made in this study showed that
perturbation theory can be used to calculate the masses of Higgs
bosons in the modified NMSSM to within 1~GeV ($\sim 1\%$).

This method can be used to diagonalise the mass matrix of a
neutralino which occurs as a result of the mixing of superpartners
of gauge bosons $W_3$ and $B$ (or $Z$ and $\gamma$) with
superpartners of neutral Higgs fields. In the basis
$(\tilde{B^0},\tilde{W^3},\tilde{H_1^0},\tilde{H_2^0},\tilde{Y})$,
this mass matrix has the following form:
\begin{equation}
\tilde{M}_{ij}=\begin{pmatrix} M_1& 0 & -A' & B' & 0\\ 0 & M_2& C'
& -D'& 0\\ -A'& C'&  0  & \mu_{\text{eff}} &
\dfrac{\lambda}{\sqrt{2}}v\sin\beta\\ B' &-D'& \mu_{\text{eff}} &
0 & \dfrac{\lambda}{\sqrt{2}}v\cos\beta\\ 0 & 0 &
\dfrac{\lambda}{\sqrt{2}}v\sin\beta &
\dfrac{\lambda}{\sqrt{2}}v\cos\beta & \mu'
\end{pmatrix} \label{19}
\end{equation}

The fragment of the $(4\times 4)$ matrix which includes the first
four columns and for rows is the same as the mass matrix of a
neutralino in the MSSM with $A'$, $B'$, $C'$, and $D'$ given by
\begin{align*}
& A'=M_Z\cos\beta\sin\theta_W, & \quad &
B'=M_Z\sin\beta\sin\theta_W,\\ & C'=M_Z\cos\beta\cos\theta_W, & &
D'=M_Z\sin\beta\cos\theta_W.
\end{align*}
Using the unitary transformation $U$:
\[
U=\begin{pmatrix} 1 & 0 & 0 & 0 & 0 \\ 0 & 1 & 0 & 0 & 0 \\ 0 & 0
& \dfrac{1}{\sqrt{2}} & \dfrac{1}{\sqrt{2}} & 0 \\ 0 & 0
&-\dfrac{1}{\sqrt{2}} & \dfrac{1}{\sqrt{2}} & 0 \\ 0 & 0 & 0 & 0 &
1
\end{pmatrix}
\]
the matrix (\ref{19}) can be reduced to the form (\ref{18})
$\tilde{M}'=U\tilde{M}U^+$ and then using the ratios
$(M_Z/\mu_{\text{eff}})^2$ and $(M_Z/\mu')^2$ as the small
parameters in the first order of perturbation theory for the
spectrum of supersymmetric particles we obtain
\begin{eqnarray}
m_{\tilde{\chi}_1}&=&M_1+\frac{(A'-B')^2}{2(M_1-\mu_{\text{eff}})}+
\frac{(A'+B')^2}{2(M_1+\mu_{\text{eff}})}\, ,\nonumber \\
m_{\tilde{\chi}_2}&=&M_2+\frac{(C'-D')^2}{2(M_2-\mu_{\text{eff}})}+
\frac{(C'+D')^2}{2(M_2+\mu_{\text{eff}})}\, ,\nonumber \\
m_{\tilde{\chi}_3}&=&\mu_{\text{eff}}+\frac{(A'-B')^2}{2(\mu_{\text{eff}}-M_1)}+
\frac{(C'-D')^2}{2(\mu_{\text{eff}}-M_2)}+ \frac{\lambda^2
v^2\sin^2(\beta+\pi/4)}{2(\mu_{\text{eff}}-\mu')}\, \label{20} \\
m_{\tilde{\chi}_4}&=&-\mu_{\text{eff}}-\frac{(A'+B')^2}{2(M_1+\mu_{\text{eff}})}-
\frac{(C'+D')^2}{2(M_2+\mu_{\text{eff}})}- \frac{\lambda^2
v^2\sin^2(\beta-\pi/4)}{2(\mu_{\text{eff}}+\mu')},\nonumber \\
m_{\tilde{\chi}_5}&=&\mu'+\frac{\lambda^2 v^2\sin^2(\beta+\pi/4)}
{2(\mu'-\mu_{\text{eff}})}+\frac{\lambda^2 v^2\sin^2(\beta-\pi/4)}
{2(\mu'+\mu_{\text{eff}})}\, . \nonumber
\end{eqnarray}
The accuracy with which $m_{\tilde{\chi}_i}$ is calculated is
slightly lower than that for the CP--even Higgs sector. This
primarily because the parameters used for the expansion when
diagonalising the neutralino mass matrix according to perturbation
theory are larger. At this point we shall not discuss the spectrum
of squarks, sleptons, and charginos in greater detail since the
analytic expressions for the masses of these particles remain the
same as in MSSM. In this case, in all the formulas we need to
replace $\mu$ with $\mu_{\text{eff}}$. We merely note that in the
principal approximation the masses of two Dirac charginos and
neutralinos $\tilde{\chi}_2$ and $\tilde{\chi}_3$ are the same:
$m_{\tilde{\chi}^\pm_1}\approx\mu_{\text{eff}}$ and
$m_{\tilde{\chi}^\pm_2}\approx M_2$.

\section{Results of the numerical analysis}

Results of a numerical analysis of the spectrum of Higgs bosons
and superpartners of observable particles in the modified NMSSM
are given in Figs.1--5 and Tables 1--3. We first need to note that
for a fixed sign of $\mu_{\text{eff}}$ there are two allowed
regions of parameter space. In one of these the mass of the
lightest Higgs boson is greater than in the MSSM (see Figs. 1a and
3a) whereas in the other it is smaller (see Figs. 1b and 3b). The
mass of the lightest calculated in the first order with respect to
perturbation theory has the following form:
\begin{equation}
m_h^2\approx V_{11}-\frac{|V_{13}|^2}{E_3^2}=M_Z^2\cos^2 2\beta
 +\frac{1}{2}\lambda^2 v^2\sin^22\beta+\Delta_{11}
-\frac{(\lambda vX_1+\Delta_{13})^2}{m_y^2+{\mu'}^2+B'\mu'}\, .
\label{21}
\end{equation}
Since the matrix element $V_{12}\sim M_Z^2$, we neglected its
contribution to $m_h$. The mass of the lightest CP--even Higgs
boson reaches its highest value when
$\mu'=-\dfrac{2\mu_{\text{eff}}}{\sin 2\beta}-A_\lambda-
\dfrac{\sqrt{2}\Delta_{13}}{\lambda v \sin 2\beta}$, when
$V_{13}=0$ (see Figs. 1a and 3a). Thus, $m_h$ is larger in that
region of parameter space where the signs of $\mu'$ and
$\mu_{\text{eff}}$ are opposed. In the limit $\mu'\to\pm\infty$
the masses of the CP--even and CP--odd Higgs bosons corresponding
to the field $Y$ become much larger than the scale of the
supersymmetry breaking. In the low energy region their
contribution to the effective interaction potential of the Higgs
fields $H_1$ and $H_2$ disappears and the mass of the lightest
Higgs boson is the same as in the MSSM. For this reason, as can be
seen from the graphs plotted in Figs. 1a,1b and 3a,3b, $m_h$
reaches a constant value when $\mu'\to\pm\infty$. The results of
the numerical calculations plotted in these figures indicate that
the two--loop corrections \cite{11} play a significant role in the
calculations of the mass of the lightest Higgs boson. In this
particular case, they reduce its mass approximately by 10~GeV.
Although the one--loop corrections increase logarithmically as the
scale of the supersymmetry breaking increases, their increase for
$M_S\gg M_Z$ is completely compensated by the log--log asymptotic
form of the two--loop corrections and $m_h$ remains almost
constant. Allowance for the loop corrections has the result that
for $\mu_{\text{eff}}<0$ the mass of the lightest Higgs boson is
greater than that in the case $\mu_{\text{eff}}>0$. This can be
attributed to the fact that $m_h$ increases with increasing mixing
in the superpartner sector of the $t$--quark ($\tilde{t}_R$ and
$\tilde{t}_L$) which is determined by the value of
$X_t=A_t+\mu_{\text{eff}}/\tan\beta$. Since $A_t<0$, the absolute
value of the mixing between $\tilde{t}_R$ and $\tilde{t}_L$ is
greater for $\mu_{\text{eff}}<0$. It should be noted that the mass
of the lightest Higgs boson is almost independent of $A$ and $m_0$
because of the weak dependence of the squark mass on the
corresponding fundamental parameters (see Tables 1 and 2).

Since that part of the parameter space in which $\mu_{\text{eff}}$
and $\mu'$ have the same sign is almost eliminated by the existing
experimental data, it is most interesting to study the spectrum of
Higgs bosons in the region where the mass of the lightest Higgs
boson is greater than that in the minimal supersymmetric model. In
this particular region of parameter space the bilinear soft
supersymmetry breaking constants and $\mu_{\text{eff}}$ have
opposite signs and near the maximum of $m_h$ the parameter is
$\mu'\sim -2\mu_{\text{eff}}/\sin 2\beta$
($|\mu'|>|\mu_{\text{eff}}|\sim M_S$). For this reason the
heaviest particle in the modified NMSSM spectrum is the CP--even
Higgs boson which corresponds to the neutral field $Y$ sice its
mass in the principal approximation with respect to perturbation
theory is $M_S^2\approx E_3^2>{\mu'}^2$ and is substantially
larger than the scale of the supersymmetry breaking. It can be
seen from Figs. 2a and 4a that the mass of the other heavy
CP--even Higgs boson ($m_H$) is almost independent of $\mu'$ since
$m_S^2\gg m_H^2$. However, the spectrum of the CP--odd Higgs
sector is determined to a considerable by the choice of
fundamental parameters. As $\mu'$ increases, the mass of the
CP--odd $Y$ increases and the latter becomes one of the heaviest
particles. For values of $\mu'\sim B'$ the mass of the lightest
CP--odd Higgs boson $m_{A_2}^2\approx m_B^2$ is very low (see
Figs. 2a, 4a), which leads to the appearance of a constraint on
$\mu'$. Nevertheless, with this choice of fundamental parameters
this Higgs boson is negligibly involved in electromagnetic and
weak interactions since the main contribution to its wave function
is made by the CP--odd component of the field $Y$. Thus, even when
its mass is relatively low, it is extremely difficult to detect
this particle experimentally. the heaviest fermion in this model
is the superpartner of the field $Y$. Its mass
$m_{\tilde{\chi}_5}$ is proportional to $\mu'$ (see (\ref{20})).
The spectrum of remaining neutralinos, charginos, squarks, and
sleptons does not depend on the choice of $\mu'$.

Since the dependence of the soft supersymmetry breaking parameters
on $A$ disappears in the strong Yukawa coupling regime on the
electroweak scale, the spectrum of superpartners of the observable
particles and also $\mu_{\text{eff}}$ and $B$ whose numerical
values are determined by solving the system of Eqs.(\ref{12}),
vary weakly when the trilinear interaction constant of the scalar
fields varies between $-M_{1/2}$ and $M_{1/2}$. Despite this, the
dependence of the Higgs boson spectrum on $A$ is conserved. This
is mainly because the bilinear interaction constant of the neutral
scalar fields $B'$ is proportional to $A$. Using the relations
(\ref{15}), we obtain
\[
B'(t_0)=\frac{1}{\zeta(t_0)}B(t_0)+\left[\left(\sigma_1(t_0)
\frac{\mu_0}{\mu'_0}-\frac{\sigma(t_0)}{\zeta(t_0)}\right)x+
\left(\omega_1(t_0)\frac{\mu_0}{\mu'_0}-\frac{\omega(t_0)}{\zeta(t_0)}\right)
\right]M_{1/2},
\]
where $x=A/M_{1/2}$. As $x$ increases for $\mu_{\text{eff}}<0$
($\mu_{\text{eff}}>0$) the bilinear interaction constant decreases
(increases) in absolute value and conversely. As $|B'|$ decreases,
the masses of the CP--even and CP--odd states corresponding to the
neutral field $Y$ converge. At the same time, an increase in the
absolute value of $B'$ leads to a decrease in $m_{A_2}^2$ which
disappears when $B'\sim\mu'$. The dependence of the Higgs boson
spectrum of the parameter $A$ for $m_0=0$ is studied in Figs. 2b
and 4b. The parameter $\mu'$ in this particular case is selected
so that the mass of the lightest Higgs boson coincides with the
upper bound on $m_h$ for $A=0$.

Although in some cases we assumed $m_0=0$ when analysing the
modified NMSSM, this limit is unacceptable from the physical point
of view since in this case the lightest (and consequently stable)
supersymmetric particle is the superpartner of the right
$\tau$--lepton which contradicts existing astrophysical
observations. However, as $m_0$ increases the mass of the
superpartner of the right $\tau$--lepton increases and even for
comparatively low values of $m_0/M_{1/2}$ the lightest particle in
the spectrum of superpartners of observable particles becomes the
neutralino. The results of the numerical calculations presented in
Tables 1 and 2 can be used to assess the influence of the
fundamental constants $A$, $m_0$, and $M_{1/2}$ on the
superpartner spectrum of the $t$--quark, gluinos, neutralinos,
charginos, and Higgs bosons. For each set of parameters listed
above we give the values of the upper bound on the mass of the
lightest Higgs boson. calculated in the one--loop and two--loop
approximations and also the corresponding $\mu_{\text{eff}}$,
$B_0$, $y$, and $\mu'$ for which $V_{13}=0$. It can be seen from
the data presented in Table 1 that the qualitative pattern of the
spectrum remains unchanged if the parameters $A$ and $m_0$ vary
within reasonable limits. It should also be noted that as $m_0^2$
increases, the masses of squarks, sleptons, Higgs bosons, and also
heavy charginos and neutralinos increases whereas the spectrum of
the lightest particles remains unchanged. The mass of a charged
Higgs boson which has not been mentioned before is almost
independent of $A$ and $\mu'$ and numerically similar results are
obtained for the mass of a charged Higgs boson $m_{H^\pm}$.

In the present paper we have made a detailed study of the
superpartner and Higgs boson spectrum for initial values of the
Yukawa constants $h_t^2(0)=\lambda^2(0)=10$ corresponding to the
scenario of the infrared quasi-fixed point in the NMSSM. The
results of the numerical calculations presented in Tables 1 and 2
indicate that for $m_t(M_t^{\text{pole}})$ and $m_3\le
2\text{~TeV}$ the mass of the lightest Higgs boson does not exceed
127~GeV. Other data presented in Table 3 indicate that the
distinguishing features of the supersymmetric particle spectrum
are conserved for $h_t^2(0)\gg\lambda^2(0)$ and
$h_t^2(0)\ll\lambda^2(0)$ as along as the Yukawa constants on the
grand unification scale are substantially larger than the gauge
constants. Nevertheless, the upper bound on the mass of the
lightest Higgs boson, the value of $\tan\beta$, and the particle
masses calculated for $\mu'=-\dfrac{2\mu_{\text{eff}}}{\sin
2\beta}-A_\lambda- \dfrac{\sqrt{2}\Delta_{13}}{\lambda v \sin
2\beta}$, when $V_{13}=0$ vary as a function of the choice of
$h_t^2(0)$ and $\lambda^2(0)$. Nevertheless, as $\lambda^2(0)$
decreases from 10 to 2, the upper bound on $m_h$ for
$M_3=1\text{~TeV}$ drops from 128 to 113~GeV (see Table 3). Thus,
at the concluding stage of the analysis of the modified NMSSM for
each fixed $\tan\beta$ we selected the Yukawa constant
$\lambda(t_0)$ so that $m_h$ reached its highest value on
condition that perturbation theory can be applied as far as the
grand unification scale. The dependence $m_h(\tan\beta)$ thus
obtained is plotted in Fig. 5 where we also plotted the upper
bound $m_h$ in the MSSM as a function of $\tan\beta$. As was to be
expected, the two bounds on the mass of the lightest Higgs boson
are almost the same for large $\tan\beta$ when the term
$\dfrac{\lambda^2 v^2}{2}\sin^2 2\beta$ in Eq.(\ref{5}) tends to
zero. The curve $m_h(\tan\beta)$ in the NMSSM reaches its maximum
when $\tan\beta\sim 2.5$ which corresponds to the strong Yukawa
coupling regime. Both bounds on the mass of the lightest Higgs
boson were obtained for $M_3\le 2\text{~TeV}$. By varying the
scale of supersymmetry breaking we can show that the mass of the
lightest Higgs boson in the NMSSM does not exceed $130.5\pm
3.5\text{~GeV}$. The indeterminacy observed in calculations of the
upper bound on $m_h$ is mainly attributable to the experimental
error with which the mass of a $t$--quark is measured.

\section{Conclusions}

In the nonminimal supersymmetric model the mass of the lightest
Higgs boson reaches its highest value in the strong Yukawa
coupling regime when all the solutions of the renormalisation
group equations are grouped near the infrared quasi-fixed point.
However, in this region of the parameter space using the NMSSM
with a minimal set of fundamental parameters it is not possible to
obtain a self-consistent solution which on the one hand would give
a spectrum with heavy supersymmetric particles and on the other
could give a mass of the lightest Higgs boson greater than that in
the MSSM. In order to find such a solution, we need to modify the
nonminimal supersymmetric model. In the present paper we studied
the spectrum of superpartners and Higgs bosons using a very simple
expansion of he NMSSM which can give a self-consistent solution in
the strong Yukawa coupling regime. Although the parameter space of
this model is expanded substantially, the theory does not lose its
predictive capacity.

The mass matrix of the CP--even Higgs sector in the modified NMSSM
has a hierarchical structure which means that it can be
diagonalised using a method of calculating the spectrum of Higgs
bosons proposed earlier, which is based on the ordinary
perturbation theory of quantum mechanics. This method can be used
to calculate the mass of Higgs bosons to within 1~GeV ($\sim
1\%$). In this case the mass of the lightest Higgs boson near the
infrared quasi-fixed point for
$m_t(M_t^{\text{pole}})=165\text{~GeV}$ and $M_3\le 2\text{~TeV}$
does not exceed 127~GeV. By varying the ratio of the Yukawa
constants on the grand unification scale, we can show that $m_h\le
130.5\pm 3.5\text{~GeV}$ where the indeterminacy observed when
calculating the upper bound on $m_h$ is mainly attributable to the
experimental error with which the mass of the $t$--quark is
measured. The heaviest particle in the region of the parameter
space of interest is the CP--even Higgs boson corresponding to the
neutral field $Y$.

In the present study we used the same method of diagonalising the
mass matrices to calculate and study neutralino masses. As a
result we showed that the heaviest fermion in the dominant region
of parameter space is $\tilde{Y}$, the superpartner of the neutral
scalar field $Y$. For values of $m_0^2\le M_{1/2}^2$ gluinos,
squarks, heavy CP--even and CP--odd Higgs bosons are substantially
heavier than sleptons, lightest charginos, and neutralinos. The
only exception is one of the CP--odd Higgs bosons whose mass
varies substantially depending on the choice of parameters of the
model. However, even if it is relatively low, for example, of the
order of $M_Z$, there are certain problems involved in recording
it experimentally since the main contribution to its wave function
is made by the CP--odd component of the field $Y$.

The upper bound on the mass of the lightest Higgs boson in the
nonminimal supersymmetric model  was also studied on recent
publications \cite{25} and \cite{62}. The predictions obtained in
these studies are $5-6\text{~GeV}$ higher than the bound given
above. The difference in the predictions can be attributed to the
fact that the authors of \cite{25} and \cite{62} used the value of
$|X_t/M_S|=\sqrt{6}$ where
$M_S=\sqrt{m_{\tilde{t}_1}m_{\tilde{t}_2}}$ to calculate the upper
bound on $m_h$ since the mass of the lightest Higgs boson reaches
its highest value for this value of $X_t$. However, in the strong
Yukawa coupling regime in the modified NMSSM the ratio $|X_t/M_S|$
is $1.4-1.5$. Since the mass of the lightest Higgs boson increases
with increasing mixing between the $t$--quark superpartners for
$0\le|X_t/M_S|\le\sqrt{6}$, and the ratio $|X_t/M_S|$ is
considerably less than $\sqrt{6}$, the upper bound on $m_h$ in the
realistic expansion of the NMSSM is more stringent that the
absolute bound in the nonminimal supersymmetric model.

\section*{Acknowledgements}

The authors are grateful to M.I.Vysotsky, D.I.Kazakov, and
K.A.Ter-Martirosyan for stimulating questions, useful discussions,
and comments. One of the authors (R.B.N.) thanks the National
Institute of Nuclear Physics, Ferrara, Italy for their
hospitality. This work was supported by the Russian Foundation for
Basic Research (projects \#\# 98-02-17372, 98-02-17453,
00-15-96786, 00-15-96562).

\newpage

\section*{Appendix}

{\bfseries Renormalisation group equations for $\mu$, $\mu'$, $B$,
and $B'$ parameters in the modified NMSSM and their solutions.}

Side by side with the trilinear couplings and masses of scalar
particles the modified NMSSM, within which we investigate the
particle spectrum, contains the parameters $\mu$, $\mu'$, $B$, and
$B'$. An evolution of these constants is described by four
renormalisation group equations:
\begin{equation}
\begin{split}
\frac{d\mu}{dt}&=-\frac{\mu}{2}\left(2Y_\lambda+3Y_t
-3\tilde{\alpha}_2-\frac{3}{5}\tilde{\alpha}_1\right),\\
\frac{d\mu'}{dt}&=-2\mu'(Y_\lambda+Y_\varkappa),\\
\frac{dB}{dt}&=-\left(2Y_\lambda B+\sqrt{Y_\lambda Y_\varkappa}
B'\frac{\mu'}{\mu}+3Y_t A_t+ 2Y_\lambda A_\lambda-
3\tilde{\alpha}_2 M_2-\frac{3}{5}\tilde{\alpha}_1 M_1\right),\\
\frac{dB'}{dt}&=-\left(2Y_\varkappa B'+4\sqrt{Y_\lambda
Y_\varkappa} B\frac{\mu}{\mu'}+4Y_\lambda A_\lambda
\frac{\mu}{\mu'}+4Y_\varkappa A_\varkappa\right).
\end{split}
\label{A1}
\end{equation}
For $\varkappa=0$ with the minimal set of fundamental parameters
$B(0)=B'(0)=B_0$, $A_i(0)=A$, $M_i(0)=M_{1/2}$, $\mu'(0)=\mu'_0$,
and $\mu(0)=\mu_0$, one can show using a general solution of the
system of linear differential equations, that
\begin{equation}
\begin{split}
\mu(t)&=\xi(t)\mu_0,\\ \mu'(t)&=\xi_1(t)\mu'_0,\\
B(t)&=\zeta(t)B_0+\sigma(t)A+\omega(t)M_{1/2},\\
B'(t)&=B_0+\sigma_1(t)A\frac{\mu_0}{\mu'_0}+\omega_1(t)M_{1/2}\frac{\mu_0}{\mu'_0},
\end{split}
\label{A2}
\end{equation}
where the functions  $\xi(t)$, $\xi_1(t)$, $\zeta(t)$,
$\sigma(t)$, $\sigma_1(t)$, $\omega(t)$, and $\omega_1(t)$, which
determine the evolution of the fundamental parameters, mainly
depend on a choice of Yukawa constants on the grand unification
scale and do not depend on the initial values of the soft SUSY
breaking parameters, $\mu_0$, and $\mu'_0$.

\newpage

\begin{center}

{\bfseries Table 1.} Mass spectrum of superpartners of observable
particles and Higgs bosons for $\lambda^2(0)=h^2_t(0)=10$ and
$\mu_{\text{eff}}>0$ depending upon the choice of fundamental
parameters $A$, $m_0$, and $M_{1/2}$ (all parameters and masses
are given in GeV).

\vspace*{5mm}

\begin{tabular}{|c|c|c|c|c|c|c|}
\hline $m^2_0$&0&$M^2_{1/2}$&0&0&0&0\\ \hline
$A$&0&0&$-M_{1/2}$&$0.5 M_{1/2}$&0&0\\ \hline
$M_{1/2}$&-392.8&-392.8&-392.8&-392.8&-785.5&-196.4\\ \hline
$m_t(t_0)$&165&165&165&165&165&165\\ \hline
$\tan\beta$&1.883&1.883&1.883&1.883&1.883&1.883\\ \hline
$\mu_{\text{eff}}$&728.6&841.7&726.8&730.1&1361.2&380.4\\ \hline
$B_0$&-1629.1&-1935.4&-1260.0&-1813.2&-3064.4&-861.8\\ \hline
$y$&-0.00037&-0.00021&-0.00043&-0.00035&-0.00006&-0.00233\\ \hline
$\mu'(t_0)$&-1899.8&-2176.7&-1905.9&-1898.3&-3544.6&-993.1\\
\hline $\mathbf{m_h (t_0)}$&{\bf 125.0}&{\bf 125.1}&{\bf
125.0}&{\bf 125.0}& {\bf 134.9}&{\bf 114.8}\\ {\bf
(1--loop)}&&&&&&\\ \hline $\mathbf{m_h (t_0)}$&{\bf 118.4}&{\bf
118.5}&{\bf 118.4}&{\bf 118.4}& {\bf 123.2}&{\bf 111.9}\\ {\bf
(2--loop)}&&&&&&\\ \hline $M_3(1\text{~TeV})$&
1000&1000&1000&1000&2000&500 \\ \hline
$m_{\tilde{t}_1}(1\text{~TeV})$&840.6&889.7&841.1&840.3&1652.0&447.4\\
\hline
$m_{\tilde{t}_2}(1\text{~TeV})$&695.1&713.6&696.6&694.3&1366.2&371.6\\
\hline $m_H(1\text{~TeV})$&898.5&1080.5&895.4&900.3&1691.0&468.8\\
\hline $m_S(1\text{~TeV})$&
2623.4&3034.3&2452.2&2706.0&4901.7&1378.0 \\ \hline
$m_{A_1}(1\text{~TeV})$&953.9&1113.8&1245.7&925.2&1722.6&538.2\\
\hline
$m_{A_2}(1\text{~TeV})$&704.3&762.7&872.0&318.2&1366.2&302.2\\
\hline
$m_{\tilde{\chi}_1}(t_0)$&164.6&164.4&164.6&164.6&326.9&84.3\\
\hline
$m_{\tilde{\chi}_2}(t_0)$&327.8&327.6&327.8&327.8&649.4&170.1\\
\hline
$m_{\tilde{\chi}_3}(1\text{~TeV})$&755.1&870.8&753.3&756.7&1404.2&400.9\\
\hline
$|m_{\tilde{\chi}_4}(1\text{~TeV})|$&755.9&872.6&755.1&758.4&1405.0&404.3\\
\hline
$|m_{\tilde{\chi}_5}(1\text{~TeV})|$&1931.8&2212.3&1938&1930.3&3599.0&1015.4\\
\hline
$m_{\tilde{\chi}^{\pm}_1}(t_0)$&327.8&327.6&327.8&327.8&649.4&169.9\\
\hline
$m_{\tilde{\chi}^{\pm}_2}(1\text{~TeV})$&757.0&872.6&755.2&758.5&1405.2&404.5\\
\hline
\end{tabular}

\end{center}

\newpage

\begin{center}

{\bfseries Table 2.} Mass spectrum of superpartners of observable
particles and Higgs bosons for $\lambda^2(0)=h^2_t(0)=10$ and
$\mu_{\text{eff}}<0$ depending upon the choice of fundamental
parameters $A$, $m_0$, and $M_{1/2}$ (all parameters and masses
are given in GeV).

\vspace*{5mm}

\begin{tabular}{|c|c|c|c|c|c|c|c|c|}
\hline $m^2_0$&0&$M^2_{1/2}$&0&0&0&0\\ \hline
$A$&0&0&$-M_{1/2}$&$M_{1/2}$&0&0\\ \hline
$M_{1/2}$&-392.8&-392.8&-392.8&-392.8&-785.5&-196.4\\ \hline
$m_t(t_0)$&165&165&165&165&165&165\\ \hline
$\tan\beta$&1.883&1.883&1.883&1.883&1.883&1.883\\ \hline
$\mu_{\text{eff}}$&-727.8&-840.9&-726.0&-731.2&-1360.7&-378.9\\
\hline $B_0$&1008&1320.3&1366.7&647.9&2050.4&495.8\\ \hline
$y$&-0.00149&-0.001&-0.00128&-0.00177&-0.00020&-0.0112\\ \hline
$\mu'(t_0)$&1671.5&1950.6&1656.8&1690.3&3172.7&857.8\\ \hline
${\bf m_h (t_0)}$&{\bf 134.1}&{\bf 134.9}&{\bf 134.0}& {\bf
134.2}&{\bf 143.1}&{\bf 124.1}\\ {\bf (1--loop)}&&&&&&\\ \hline
${\bf m_h (t_0)}$&{\bf 124.4}&{\bf 124.8}&{\bf 124.3}&{\bf 124.5}
&{\bf 127.2}&{\bf 119.6}\\ {\bf (2--loop)}&&&&&&\\ \hline
$M_3(1\text{~TeV})$&1000&1000&1000&1000&2000&500\\ \hline
$m_{\tilde{t}_1}(1\text{~TeV})$&890.2&935.6&890.5&889.8&1682.8&507.9\\
\hline
$m_{\tilde{t}_2}(1\text{~TeV})$&630.3&652.2&632.2&628.0&1328.1&283.5\\
\hline $m_H(1\text{~TeV})$&896.2&1078.5&893.5&899.3&1689.9&464.4\\
\hline
$m_S(1\text{~TeV})$&2147.4&2565.9&2309.2&1972.3&4126.5&1097.7\\
\hline
$m_{A_1}(1\text{~TeV})$&1123.2&1219.3&931.0&1437.9&1984.8&623.1\\
\hline
$m_{A_2}(1\text{~TeV})$&857.6&1017.8&545.0&886.9&1657.5&412.8\\
\hline
$m_{\tilde{\chi}_1}(t_0)$&160.0&160.5&160.0&160.0&324.4&74.9\\
\hline
$m_{\tilde{\chi}_2}(t_0)$&311.1&313.7&311.0&311.2&639.9&141.4\\
\hline
$|m_{\tilde{\chi}_3}(1\text{~TeV})|$&753.7&896.6&751.9&757.2&1403.4&398.5\\
\hline
$m_{\tilde{\chi}_4}(1\text{~TeV})$&764.7&878.1&763.0&768.1&1410.0&416.7\\
\hline
$m_{\tilde{\chi}_5}(1\text{~TeV})$&1700.7&1983.2&1685.8&1719.6&3221.8&879.1\\
\hline
$m_{\tilde{\chi}^{\pm}_1}(t_0)$&310.7&313.4&310.7&310.8&639.8&139.4\\
\hline
$m_{\tilde{\chi}^{\pm}_2}(1\text{~TeV})$&763.3&877.0&761.6&766.7&1409.1&414.5\\
\hline
\end{tabular}

\end{center}

\newpage

\begin{center}

{\bfseries Table 3.} Mass spectrum of superpartners of observable
particles and Higgs bosons for $A=m_0=0$,  but for different
initial values $h_t^2(0)$,$\lambda^2(0)$\\ (all parameters and
masses are given in GeV).

\vspace*{5mm}

\begin{tabular}{|c|c|c|c|c|c|c|c|c|}
\hline
&\multicolumn{4}{|c|}{$\mu_{\text{eff}}<0$}&\multicolumn{4}{|c|}{$\mu_{\text{eff}}>0$}\\
\hline $\lambda^2(0)$&0&2&10&10&0&2&10&10\\ \hline
$h_t^2(0)$&10&10&10&2&10&10&10&2\\ \hline
$M_{1/2}$&-392.8&-392.8&-392.8&-392.8&-392.8&-392.8&-392.8&-392.8\\
\hline
$\tan\beta$&1.614&1.736&1.883&2.439&1.614&1.736&1.883&2.439\\
\hline
$\mu_{\text{eff}}$&-821.5&-771.4&-727.8&-641.8&822.7&772.4&728.6&642.3\\
\hline
$B_0$&471.7&622.5&1008.0&886.2&-743.1&-988.1&-1629.1&-1583.3\\
\hline
$y$&---&-0.0014&-0.0015&-0.0012&---&-0.0003&-0.0004&-0.0005\\
\hline
$\mu'(t_0$)&---&1693.9&1671.5&1749.8&---&-1941.4&-1899.8&-1943.1\\
\hline ${\bf m_h (t_0)}$&{\bf 103.5}&{\bf 123.6}&{\bf 134.1}&{\bf
137.6} &{\bf 88.1}&{\bf 112.4}&{\bf 125.0}&{\bf 131.2}\\ {\bf
(1--loop)}&&&&&&&&\\ \hline ${\bf m_h (t_0)}$&{\bf 90.3}&{\bf
113.0}&{\bf 124.4}&{\bf 127.8} &{\bf 79.7}&{\bf 105.5}&{\bf
118.4}&{\bf 123.6}\\ {\bf (2--loop)}&&&&&&&&\\ \hline
$M_3(1\text{~TeV})$&1000&1000&1000&1000&1000&1000&1000&1000\\
\hline
$m_{\tilde{t}_1}(1\text{~TeV})$&894.0&891.6&890.2&890.5&834.6&837.0&840.6&853.5\\
\hline
$m_{\tilde{t}_2}(1\text{~TeV})$&613.5&622.2&630.3&648.5&692.2&693.8&695.1&696.4\\
\hline
$m_H(1\text{~TeV})$&1033.4&961.0&896.2&758.5&1035.7&963.3&898.5&761.1\\
\hline
$m_S(1\text{~TeV})$&---&1999.8&2147.4&2187.2&---&2405.3&2623.4&2663.8\\
\hline
$m_{A_1}(1\text{~TeV})$&1029.7&1374.8&1123.2&1294.0&1031.3&1390.6&953.9&965.1\\
\hline
$m_{A_2}(1\text{~TeV})$&---&949.8&857.6&735.6&---&951.6&704.3&674.3\\
\hline
$m_{\tilde{\chi}_1}(t_0)$&160.3&160.1&160.0&159.9&164.6&164.6&164.6&164.4\\
\hline
$m_{\tilde{\chi}_2}(t_0)$&312.7&311.9&311.1&309.4&328.2&328.1&327.8&326.4\\
\hline
$|m_{\tilde{\chi}_3}(1\text{~TeV})|$&842.8&795.8&753.7&665.8&844.4&797.2&755.1&
668.1\\ \hline
$|m_{\tilde{\chi}_4}(1\text{~TeV})|$&856.4&807.8&764.7&677.1&850.6&800.9&755.9&
666.7\\ \hline
$|m_{\tilde{\chi}_5}(1\text{~TeV})|$&---&1711.2&1700.7&1790.0&---&1960.7&1931.8
&1986.5\\ \hline
$m_{\tilde{\chi}^{\pm}_1}(t_0)$&312.4&311.6&310.7&309.0&328.2&328.1&327.8
&326.4\\ \hline
$m_{\tilde{\chi}^{\pm}_2}(1\text{~TeV})$&854.2&806.0&763.3&676.7&849.5&800.4&
757.0&669.0\\ \hline
\end{tabular}

\end{center}

\newpage

\section*{Figure captions}

{\bfseries Fig.1.} The lightest Higgs boson mass in modified NMSSM
as a function of $z=\mu'/1\text{~TeV}$ for
$h_t^2(0)=\lambda^2(0)=10$, $m_0=0$, $M_3=1\text{~TeV}$, and for
$\mu_{\text{eff}}\le 0$. Thick and thin curves meet calculations
in one--loop and two--loop approximations, respectively.\\

{\bfseries Fig.2.} Mass spectrum in modified NMSSM as a function
of $z=\mu'/1\text{~TeV}$ and $x=A/M_{1/2}$ for
$h_t^2(0)=\lambda^2(0)=10$, $m_0=0$, $M_3=1\text{~TeV}$, and for
$\mu_{\text{eff}}\le 0$. Thick and thin curves give the masses of
heavy CP--even Higgs bosons. Dotted and dashed curves give the
masses of CP--odd Higgs bosons. The dashdot curve gives the mass
of the heaviest neutralino.\\

{\bfseries Fig.3.} The lightest Higgs boson mass in modified NMSSM
as a function of $z=\mu'/1\text{~TeV}$ for
$h_t^2(0)=\lambda^2(0)=10$, $m_0=0$, $M_3=1\text{~TeV}$, and for
$\mu_{\text{eff}}\ge 0$. Thick and thin curves meet calculations
in one--loop and two--loop approximations, respectively.\\

{\bfseries Fig.4.} Mass spectrum in modified NMSSM as a function
of $z=\mu'/1\text{~TeV}$ and $x=A/M_{1/2}$ for
$h_t^2(0)=\lambda^2(0)=10$, $m_0=0$, $M_3=1\text{~TeV}$, and for
$\mu_{\text{eff}}\ge 0$. Thick and thin curves give the masses of
heavy CP--even Higgs bosons. Dotted and dashed curves give the
masses of CP--odd Higgs bosons. The dashdot curve gives the mass
of the heaviest neutralino.\\

{\bfseries Fig.5.} Upper bound on the mass of the lightest Higgs
boson in the MSSM (thin curve) and in the modified NMSSM (thick
curve) as a function of $\tan\beta$ for $M_3=2\text{~TeV}$.

\newpage

\noindent
\includegraphics[width=159mm, keepaspectratio=true]{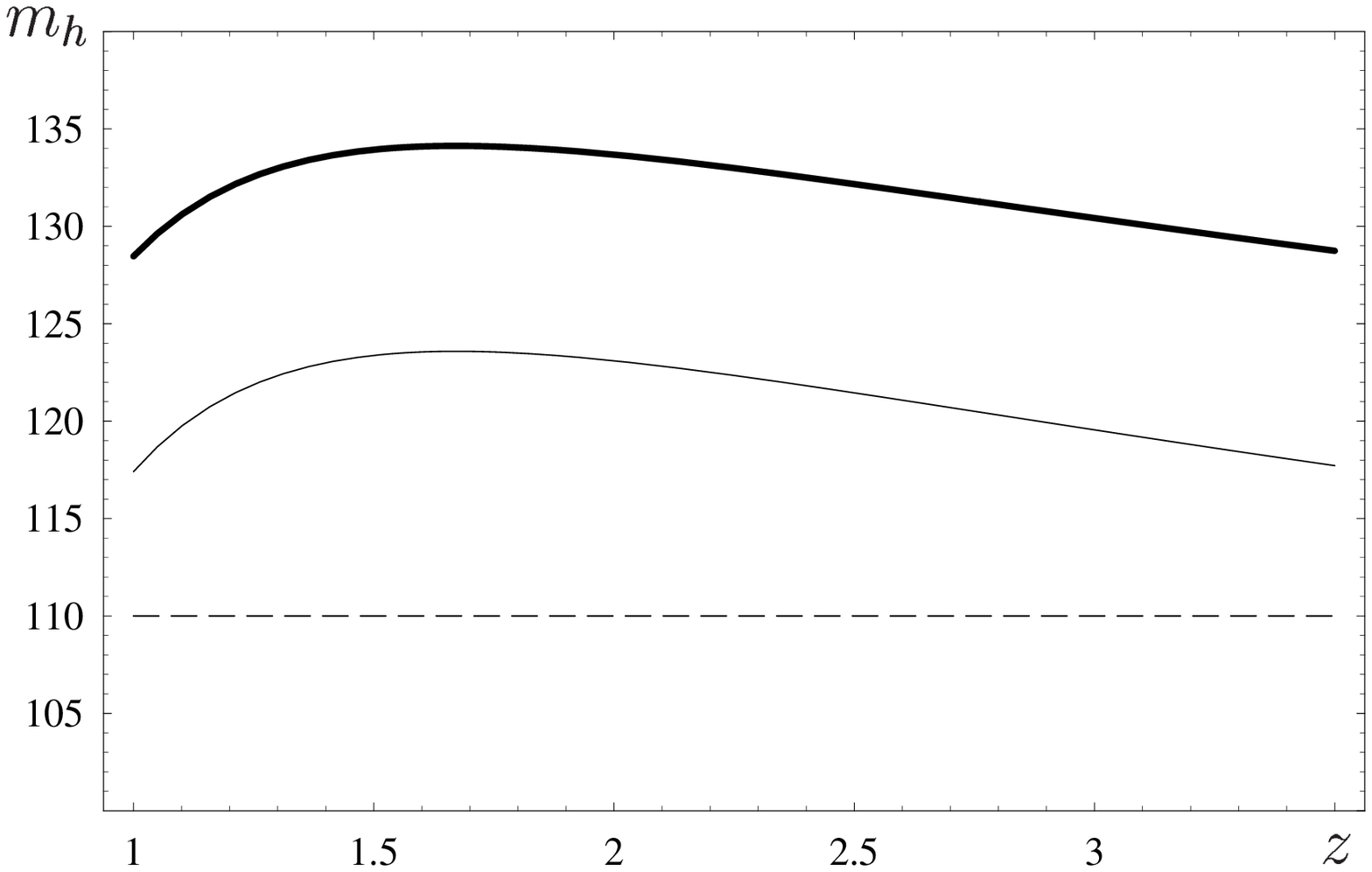}

\vspace{5mm}\hspace*{70mm}{\large\bfseries Fig.1a.}

\vspace{17mm}

\noindent
\includegraphics[width=159mm, keepaspectratio=true]{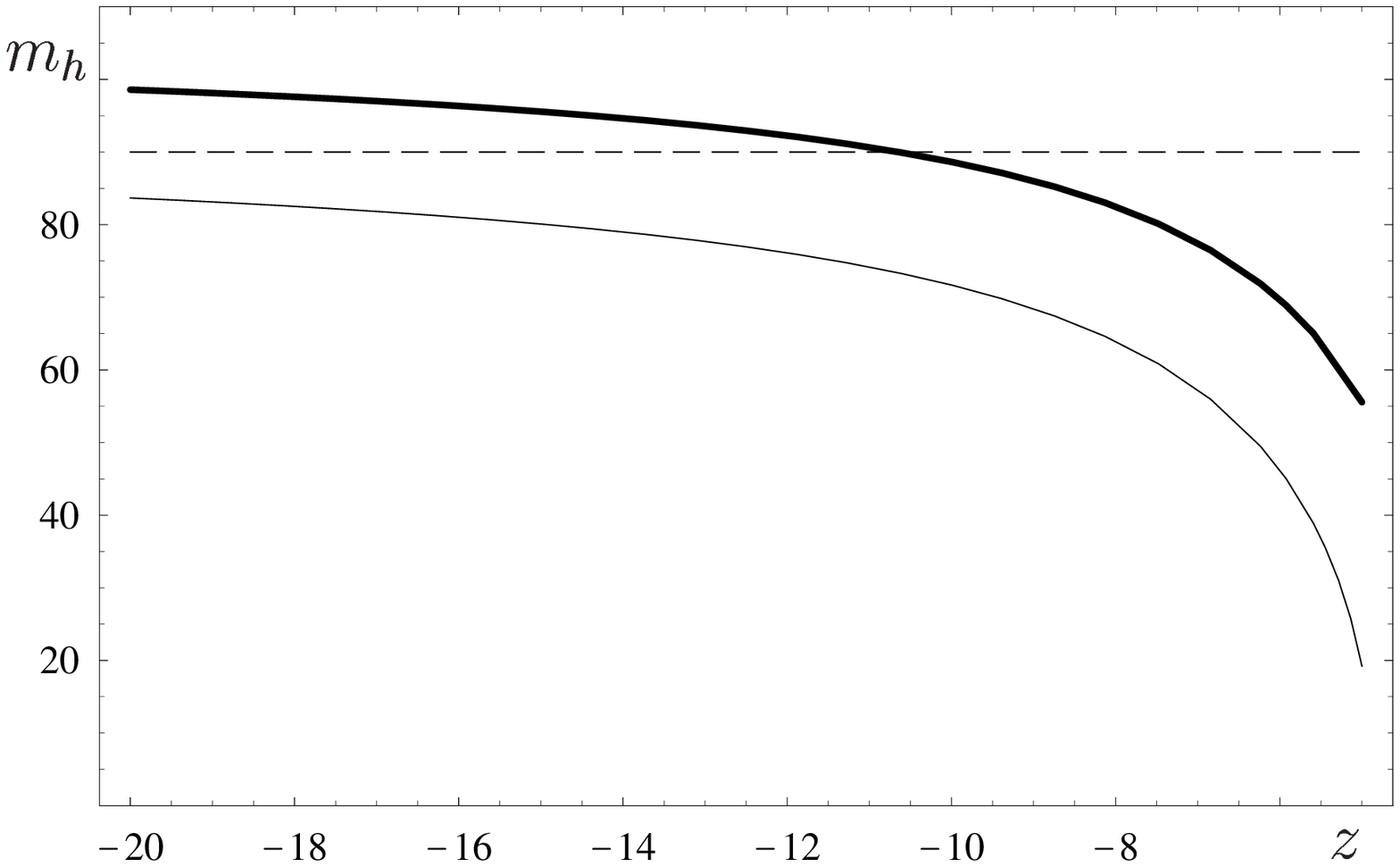}

\vspace{5mm}\hspace*{70mm}{\large\bfseries Fig.1b.}

\newpage

\noindent
\includegraphics[width=159mm, keepaspectratio=true]{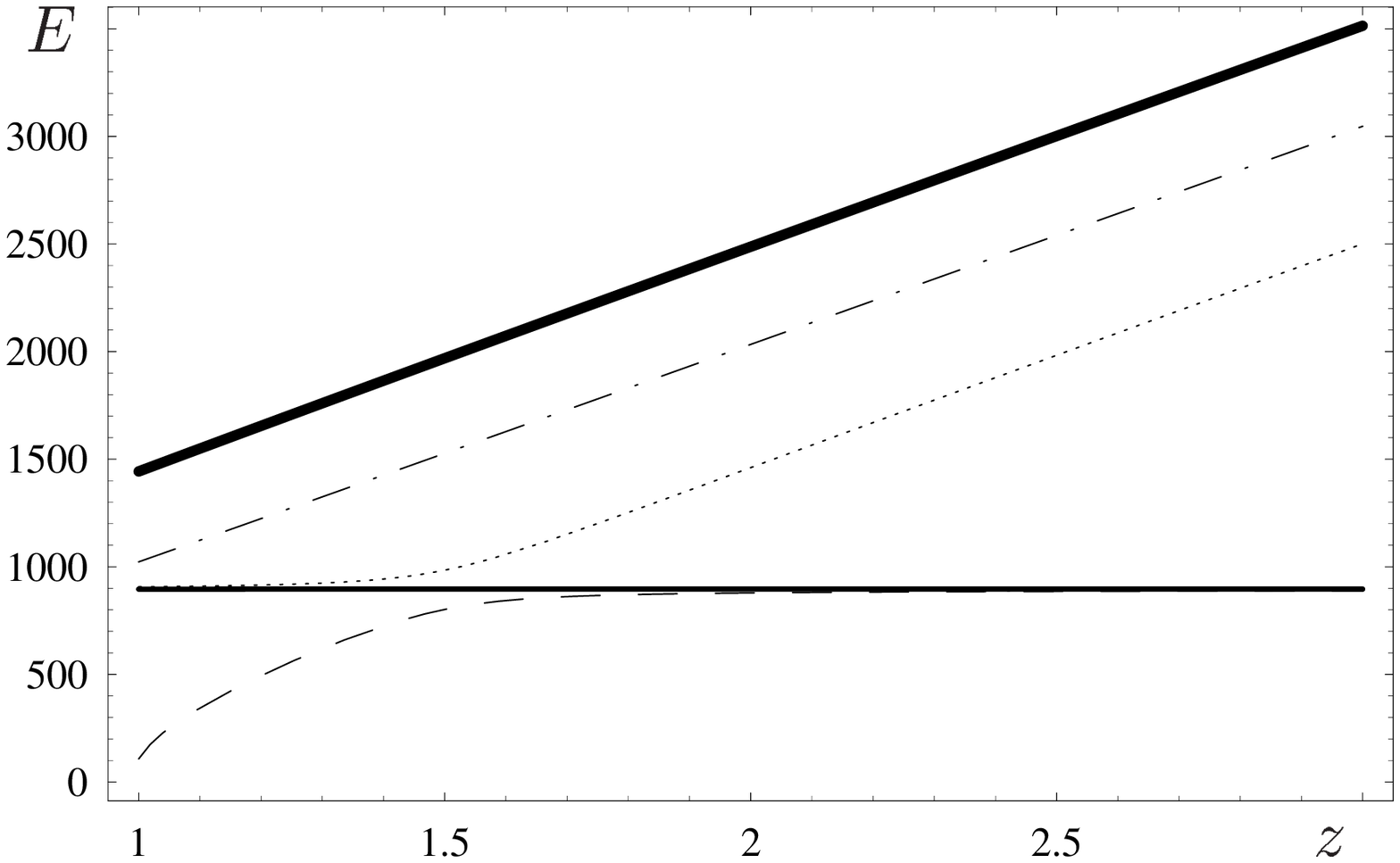}

\vspace{5mm}\hspace*{70mm}{\large\bfseries Fig.2a.}

\vspace{19mm}

\noindent
\includegraphics[width=159mm, keepaspectratio=true]{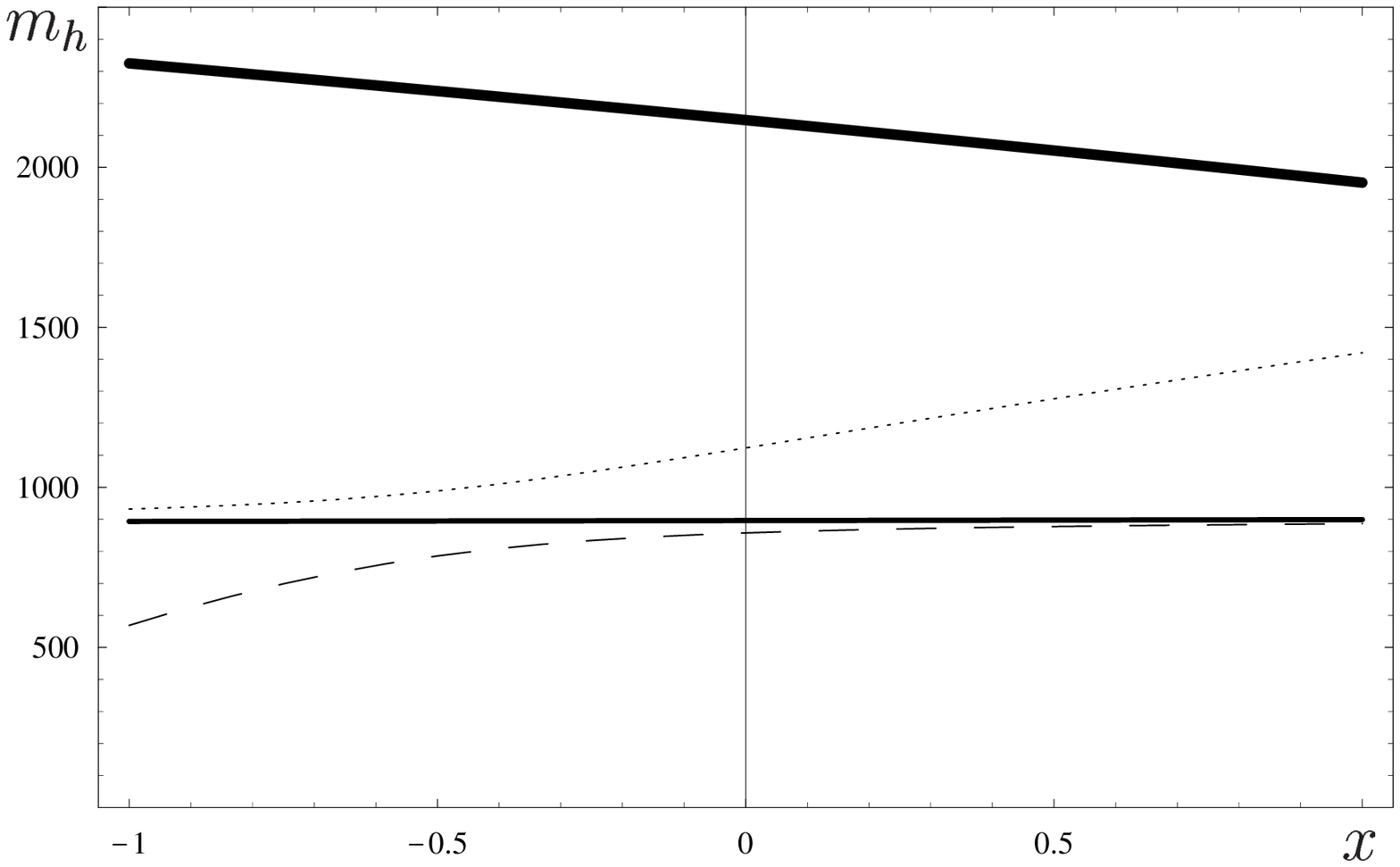}

\vspace{5mm}\hspace*{70mm}{\large\bfseries Fig.2b.}

\newpage

\noindent
\includegraphics[width=159mm, keepaspectratio=true]{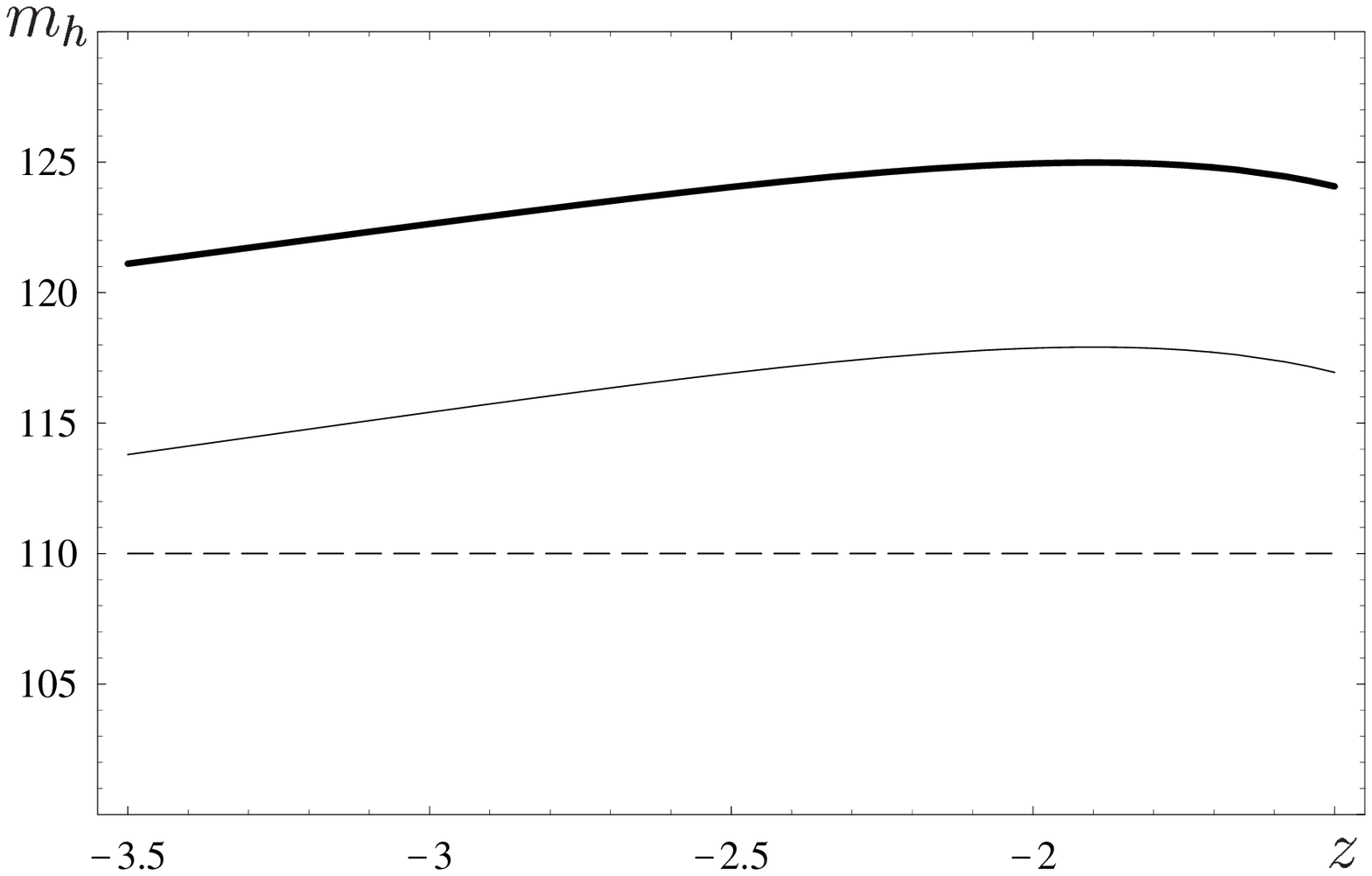}

\vspace{5mm}\hspace*{70mm}{\large\bfseries Fig.3a.}

\vspace{13mm}

\noindent
\includegraphics[width=159mm, keepaspectratio=true]{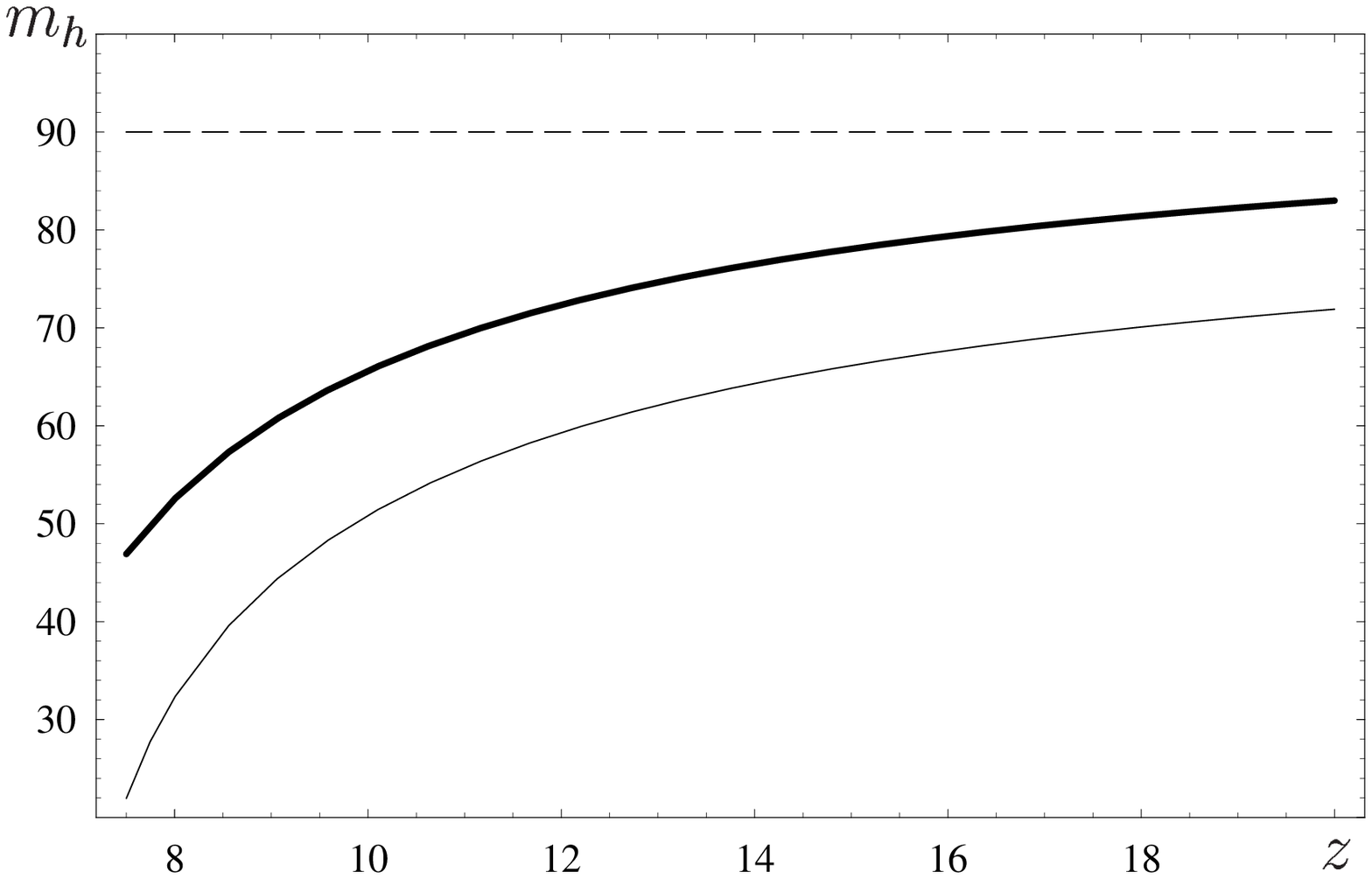}

\vspace{5mm}\hspace*{70mm}{\large\bfseries Fig.3b.}

\newpage

\noindent
\includegraphics[width=159mm, keepaspectratio=true]{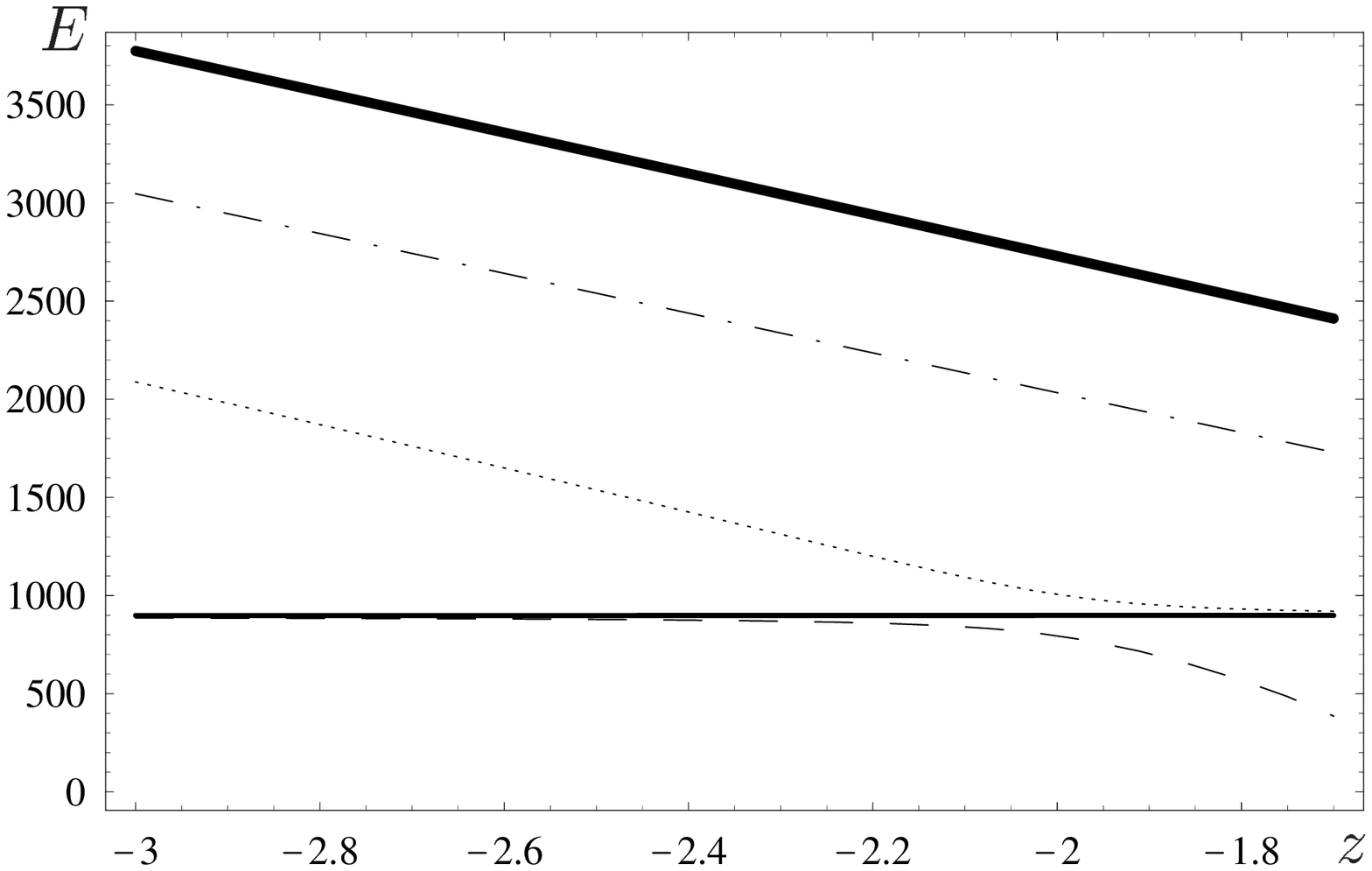}

\vspace{5mm}\hspace*{70mm}{\large\bfseries Fig.4a.}

\vspace{15mm}

\noindent
\includegraphics[width=159mm, keepaspectratio=true]{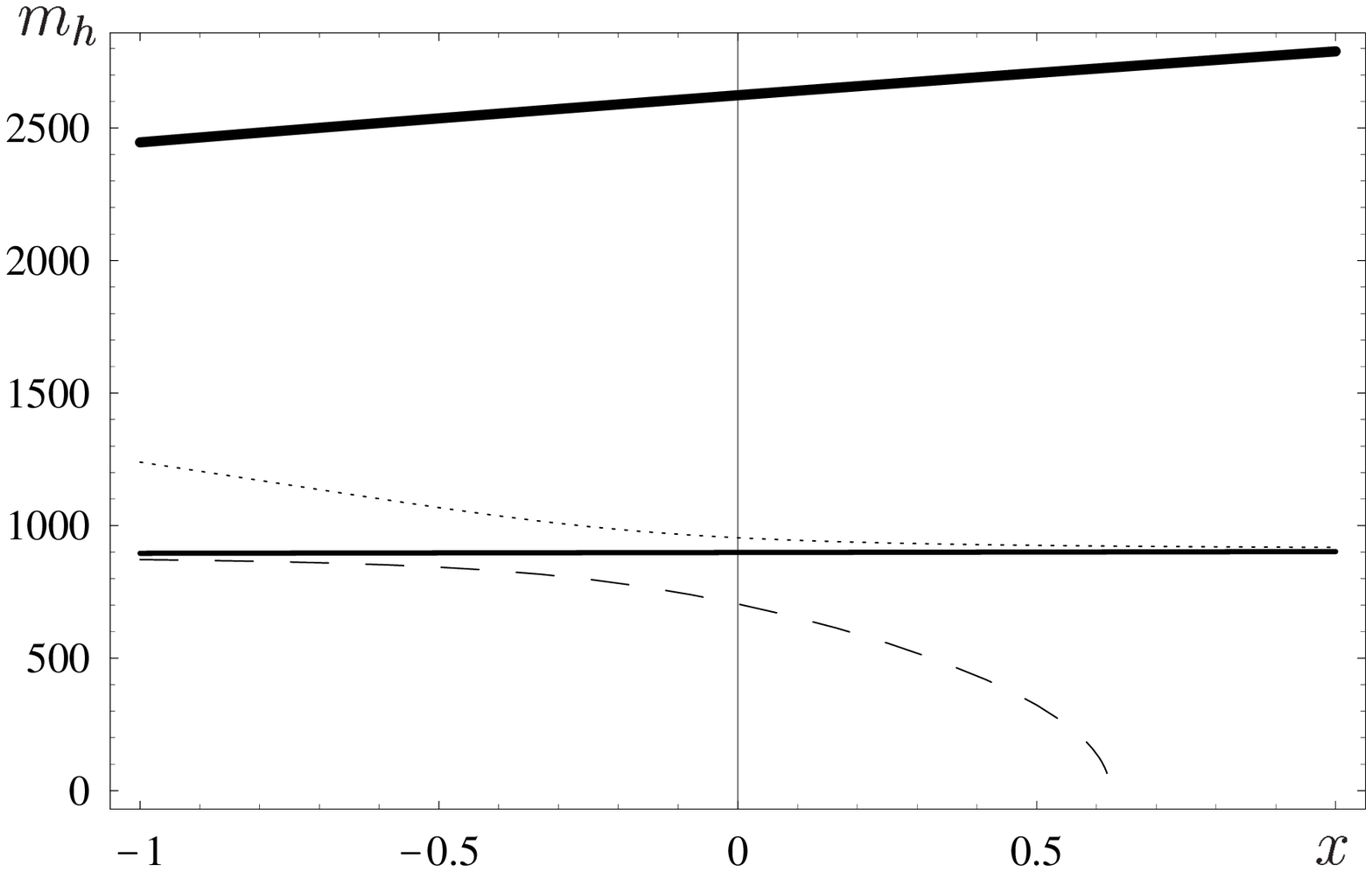}

\vspace{5mm}\hspace*{70mm}{\large\bfseries Fig.4b.}

\begin{landscape}

\vspace*{-9mm} \noindent\hspace*{3mm}
\includegraphics[width=150mm, keepaspectratio=true, angle=-90]{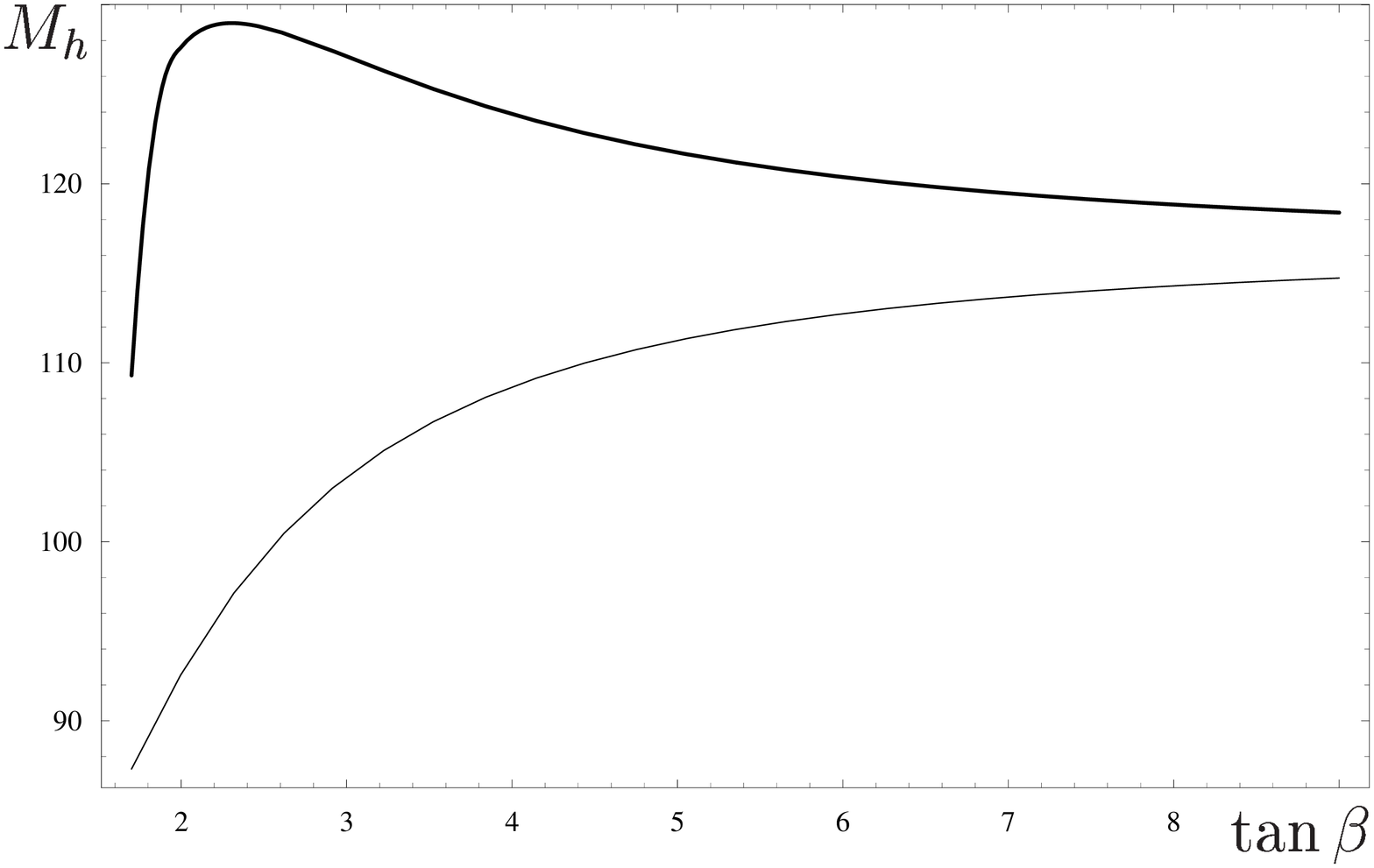}

\vspace*{-1mm}\hspace*{117.5mm}{\large\bfseries Fig.5.}

\end{landscape}

\end{document}